\documentclass[sigconf, nonacm]{acmart}

\usepackage[latin1]{inputenc}
\usepackage{epic,eepic,amsmath,latexsym,color,amsthm}
\usepackage{ifthen,graphicx}
\usepackage{times}
\usepackage[english]{babel}
\usepackage[ruled,noend,linesnumbered]{algorithm2e} 
\usepackage{subcaption}
\usepackage{breqn}
\usepackage{multirow}
\usepackage{hyperref}
\usepackage{cleveref}
\usepackage{tabto}
\usepackage{hhline}
\usepackage{enumitem}

\usepackage{balance}

\usepackage{ifthen}
\usepackage{calc}
\usepackage[normalem]{ulem} 
\definecolor{darkspringgreen}{rgb}{0.09, 0.45, 0.27}

\makeatletter   
\newcommand\ifnotfloat[1]{\@ifundefined{@captype}{#1}{}}
\newcommand\ifnotfloatelse[2]{\@ifundefined{@captype}{#1}{#2}}
\makeatother

\newboolean{withinSout}
\setboolean{withinSout}{false}

\newcommand{\ftasout}[1]{%
  \ifthenelse{\boolean{withinSout}}{
    #1
  }
  {%
    \setboolean{withinSout}{true}\sout{#1}\setboolean{withinSout}{false}%
  }%
}%

\newcommand{\ftanote}[2]{%
  \ifthenelse{\boolean{withinSout}}{}{%
    \ifnotfloat{\protect\marginpar{%
      \centering #1 \fbox{#2}}%
    }%
  }%
}

\newcommand{\annote}[3]{{\color{#3}%
		\colorbox{#3}{\bfseries\sffamily\tiny\textcolor{white}{#2}}
		\color{#3}
		$\blacktriangleright$\emph{#1}$\blacktriangleleft$}%
    \ftanote{\color{#3} \bfseries\sffamily\tiny}{#2}
}

\newcommand{\del}[1]{{\color{gray}{\ftasout{#1}}}} 

\newcommand{\alt}[1]{\annote{#1}{Alt}{magenta}}

\newcommand{\commentFT}[1]{\annote{#1}{FT}{magenta}}

\newcommand{\commentDF}[1]{\annote{#1}{DF}{darkspringgreen}}
\newcommand{\commentDB}[1]{\annote{#1}{DB}{green}}

\newcommand{\db}[1]{\commentDB{#1}}
\newcommand{\df}[1]{\commentDF{#1}}
\newcommand{\ft}[1]{\commentFT{#1}}

\renewcommand{\annote}[3]{}
\renewcommand{\del}[1]{}

\crefname{figure}{Fig.}{Fig.}
\crefname{section}{Sec.}{Sec.}

\newcommand{\institutional}{institutional\xspace}
\newcommand{\Institutional}{Institutional\xspace}
\newcommand{\infrastructure}{institutional\xspace}

\begin{document}

\newtheorem{property}{Property}
\newtheorem{assumption}{Assumption}

\newcommand{\basalt}{\textsc{Basalt}\xspace}

\newcommand{\ourmethod}{\basalt}
\newcommand{\Ourmethod}{\basalt}
\newcommand{\OurMethod}{\basalt}
\newcommand{\ourm}{\basalt}

\newcommand{\Pull}{{\sc Pull}}
\newcommand{\Push}{{\sc Push}}
\newcommand{\seed}[1]{\ensuremath{\mathrm{seed}[#1]}}
\newcommand{\view}[1]{\ensuremath{\mathrm{view}[#1]}}
\newcommand{\counter}[1]{\ensuremath{\mathrm{hits}[#1]}}
\newcommand{\rank}[2]{\ensuremath{\mathrm{rank}_{#2}(#1)}}


\title{
	\basalt: A Rock-Solid Foundation for Epidemic Consensus Algorithms
	in Very Large, Very Open Networks
}


\author{Alex Auvolat,
			Y\'{e}rom-David Bromberg, 
			Davide Frey, 
			François Taïani
}
\affiliation{%
  \institution{Univ Rennes, Inria, CNRS, IRISA, Rennes, France}
}
\email{{alex.auvolat,davide.frey}@inria.fr, {david.bromberg,francois.taiani}@irisa.fr}


\begin{abstract}
	Recent works have proposed new Byzantine consensus algorithms for blockchains
	based on epidemics, a design which enables highly-scalable performance
	at a low cost.
	These methods however critically depend on a secure random peer sampling
	service: a service that provides a stream of random network nodes
	where no attacking entity can become over-represented.
	To ensure this security property, current epidemic platforms use a Proof-of-Stake
	system to select peer samples. However such a system limits the openness
	of the system as only nodes with significant stake can participate in
	the consensus, leading to an oligopoly situation.
	Moreover, this design introduces a complex interdependency between the
	consensus algorithm and the cryptocurrency built upon it.

	In this paper, we propose a radically different security design for the
	peer sampling service, based on the distribution of IP addresses
	to prevent Sybil attacks. We propose a new algorithm, \basalt,
	that implements our design using a stubborn chaotic search to counter
	attackers' attempts at becoming over-represented.
	We show in theory and using Monte Carlo
	simulations that \basalt provides samples which are extremely close
	to the optimal distribution even in adversarial scenarios such as tentative Eclipse attacks.
	Live experiments on a production cryptocurrency
	platform confirm that the samples obtained using \basalt
	are equitably distributed amongst nodes, allowing for a system which is both open
	and where no single entity can gain excessive power.
\end{abstract}

\maketitle

\section{Introduction}


Blockchain-based systems, such as cryptocurrencies~\cite{nakamoto2009bitcoin} and smart contract platforms~\cite{ethereum_url},
are said to be \emph{Byzantine Fault Tolerant} (BFT for short),
i.e.~they are able to resist to attacks from malicious participants (called \emph{Byzantine} nodes), 
making it arbitrarily hard for instance for an attacker to forge false transactions or revoke already committed transactions.
In particular, decision power over the blockchain's state
must be spread over various network participants
in order to prevent an attacker from obtaining full control over the system.

The breakthrough made by Bitcoin~\cite{nakamoto2009bitcoin} allowed for Byzantine fault-tolerance
to be achieved in a truly open network, using a Proof-of-Work system
that requires participants to solve computationally intensive crypto-puzzles.
The difficulty of these crypto-puzzles limits the influence of individual nodes,
but encourage a race for computing power,
with Bitcoin reported to consume as much  electricity as Austria in 2020~\cite{electric_consumption_bitcoin}.
Moreover, the throughput and latency of Proof-of-Work (PoW) systems are
restricted by the time between blocks,
which must be long enough to ensure security.

\paragraph{Epidemic BFT algorithms}
A particularly interesting area of research in alleviating these issues with Proof-of-Work
consists in a new family of BFT algorithms~\cite{guerraoui2019consensus,guerraoui2019scalable,rocket2018snowflake}
that exploits epidemic mechanisms to provide large-scale protection against Byzantine behaviour.
Epidemic algorithms allow for extremely fast dissemination of information in very large networks
by means of stochastic peer-to-peer exchanges~\cite{demers1988epidemic,kermarrec_probabilistic_2003}.
Epidemic BFT algorithms exploit this property by repeatedly sampling small sets of random peers in the network,
which they then use to estimate the overall system's state,
and ensure coordination and agreement between correct (i.e. non-Byzantine) nodes.

Epidemic BFT approaches critically depend on the availability of \emph{good} network samples,
in the sense that the proportion of Byzantine nodes in a sample should be kept as low as possible,
and sampled nodes should be as varied as possible.
Providing such samples is the role of a so-called \emph{Byzantine-tolerant}, or \emph{secure}, \emph{random peer sampling} (RPS) service.
When such a service is available,
these algorithms
have the potential to yield much higher throughput than PoW systems at a fraction of the cost~\cite{rocket2018snowflake}.

\paragraph{Secure random peer sampling}
Unfortunately, classical RPS algorithms~\cite{jelasity2007gps,nedelec_adaptive_2018,voulgaris2005cyclon} are not resilient to malicious behavior:
Byzantine nodes can easily disrupt their execution by flooding honest nodes with Byzantine identifiers.
Left unchecked, this strategy has the potential to isolate honest nodes in a so-called \emph{Eclipse attack}~\cite{singh2006eclipse,heilman2015eclipse},
or to partition the system.
Moreover, a scheme where peers are sampled with uniform probability
is vulnerable to so-called \emph{Sybil attacks}~\cite{douceur2002sybil}
where a malicious entity creates arbitrarily many network node identifiers that it controls,
thus gaining unlimited influence on the network.

Current deployments of epidemic BFT algorithms,
such as the AVA cryptocurrency platform~\cite{avalabs_url},
rely on a Proof-of-Stake mechanism to ensure that nodes are sampled in a secure way,
i.e.~that the cost for an attacker of biasing samples in their favor is very high.
However, Proof-of-Stake has several known limitations~\cite{zhang2017rem}.
In essence, Proof-of-Stake consists in building an abstraction of a closed (permissioned)
system, where system membership can however evolve dynamically according to
the various parties' economic investments (in the form of token staking).
We argue that such an abstraction is too restrictive and in fact not required.
Particularly in the case of epidemic BFT algorithms,
we show that the required Byzantine-tolerant random peer sampling service
can be implemented directly in a much more open fashion,
without resorting to Proof-of-Stake to ensure security.

\paragraph{Content of this paper}
In this paper, we revisit the problem of secure peer sampling in
large-scale decentralized systems, and propose \ourmethod{}, a novel
Byzantine-tolerant random peer sampling algorithm.
\ourmethod{} exhibits close to optimal Byzantine fault tolerance, thus
significantly improving on the
state-of-the-art~\cite{bortnikov2009brahms,jesi2010secure}. \ourmethod
is designed to operate in Internet-scale permissionless systems while
resisting to Eclipse and Sybil attacks.
At the core of \ourmethod lies what
we have termed a \emph{stubborn chaotic search}, a greedy epidemic
procedure~\cite{voulgaris_vicinity} towards random nodes that are
implicitly defined in a way that makes it extremely hard for malicious
nodes to manipulate the decisions of correct ones.
This procedure is parametrized by a target distribution on nodes
based on their IP addresses, which we define to defend
against Sybil attacks by institutions that own
large contiguous portions of the IP address space.

We comprehensively analyze \ourmethod{} under a theoretical model
based on the \emph{power} $f$ of the attack, which captures the
(ideal) probability of sampling malicious nodes as defined by the target distribution.
We show that \ourmethod{} provides samples in which the proportion of malicious nodes is
very close to $f$, its theoretical optimum,
and that $f$ is acceptably small in several real-world scenarios including \institutional attacks
and botnet attacks.
We complement our theoretical model with Monte Carlo simulations that confirm our analysis.
Finally, we demonstrate the feasibility and concrete benefits of our technique by deploying \ourmethod
within a live cryptocurrency network using a prototype
implementation of \ourmethod{} for AvalancheGo~\cite{avalanchego_url},
the reference engine powering the AVA cryptocurrency
network~\cite{avalabs_url, rocket2018snowflake}.
Our experiments on the AVA network confirm that the samples obtained using \basalt
are equitably distributed amongst nodes, allowing for a system which is both open
and where no single entity can gain excessive power.
Our prototype is publicly available, fully functional, and compatible with
the existing AVA network without requiring any protocol
changes.

\section{Problem Statement}
\label{sec:problem-statement}

A random peer sampling (RPS) service can be defined as a service that produces
a continuous stream $(p_i)_{i\ge 0}$ of random nodes selected in the network.
As stated above, a secure random peer sampling service is faced with the double
task of \emph{(i)} ensuring the largest possible diversity of peers
in the stream $(p_i)_{i\ge 0}$, while \emph{(ii)} limiting as much as
possible the appearance of malicious nodes in $(p_i)_{i\ge 0}$.
\db{twice times $(p_i)$ looks weird in the same sentence}

\subsection{System Model}

We assume a very large system composed of nodes that can either be honest (a.k.a.~correct) or malicious (a.k.a.~Byzantine).
Byzantine nodes may deviate arbitrarily from the prescribed protocol
in order to manipulate the decisions taken by correct nodes,
for instance to isolate correct nodes or
to increase malicious nodes' representation in the peer sampler's output.
We write $Q$ the number of correct nodes in the system.

We consider a communication network where any node can send a message
to any other node, and assume that more than a fixed fraction of the
messages sent to a node by other non-malicious nodes arrive within a
certain delay. Byzantine nodes
may collude (share information, coordinate their behaviors), and may
send arbitrary messages to an arbitrarily large number of correct
nodes per time unit.  They cannot however block completely the
communication between two correct nodes, or read in the local memory
of correct nodes. 

Nodes are granted each a unique identifier, which we assume to be their IP address.
We will use the same notation to refer to a node and to its identifier.
We assume that Byzantine nodes may not spoof the IP addresses
of other nodes, which can be prevented using 
a handshaking mechanism~\cite{ehrenkranz2009state}.

\subsection{Sybil Attacks}
\label{sec:sybil_attack_def}

Random peer sampling is often considered under the assumption of
a \emph{closed}, or \emph{permissioned} system
(e.g.~\cite{bortnikov2009brahms,jelasity2007gps}),
where the whole set of nodes is known and the proportion of malicious
nodes is equal to (or bounded by) a small fixed fraction $\varphi$.
In such a situation, a perfect random peer sampler could be defined
as one that samples all nodes uniformly, thus returning
a fraction $\varphi$ of malicious nodes in the samples it produces.

This assumption is however not adapted to an open network such as the
public Internet, which is more akin to a \emph{permissionless} (open) system.
In such a setting, an attacker may control nodes with many times more
IP addresses than there are correct nodes, which may then be used to perform a
Sybil attack, leading to an increased influence of the attacker in the peer sample's output.
In particular, a RPS that samples peers uniformly based on their IP
addresses is particularly vulnerable to such attacks.

Drawing on the classification from~\cite{heilman2015eclipse},
we will consider two paradigmatic scenarios where an attacker attempts a Sybil attack
using many IP addresses:

\begin{enumerate}[label=(\roman*)]
	\item \textbf{\Institutional attacks}, launched by an institution or an organization
		that owns large IP address blocks; and
	\item \textbf{Botnet attacks}, where many infected machines are controlled
		by an attacker.
\end{enumerate}

The crucial difference between these two attacks is that in an
\institutional attack the attacker may control many IP addresses
located in a limited number of continuous address blocks, whereas in a
botnet attack the attacker may control a smaller number of addresses
in the whole IP address space.  These properties allow us to implement
efficient defenses by biasing our sample selection to limit the
influence of any given entity (Section~\ref{sec:sybil_biased_sampling}).
\commentFT{VLDB: Copied from USENIX Sec. response}%
From a practical perspective, these two
attacks represent the two extremes of a continuous spectrum,
as most actual attacks will usually fall somewhere in the
middle, a point we return to in our evaluation.

We do not consider network-level attacks such as BGP hijacks in our attack model,
however we discuss these attacks and potential defenses in Section~\ref{sec:network-attack}.

\section{The \OurMethod{} Algorithm}
\label{sec:algorithm}

\Ourmethod{} leverages three main components. First it employs a novel
sampling approach, termed \emph{stubborn chaotic search}, that
exploits  ranking functions to define a dynamic target random graph
(i.e.  a set of $v$ target neighbors for each node) that cannot be
controlled by Byzantine nodes. Second, it adopts a \emph{hit-counter}
mechanism that favors the exploration of new peers even in the
presence of Byzantine nodes that flood the network with their
identities. Finally, it incorporates  hierarchical ranking functions
that ensure that nodes sample their peers from a variety of address
prefixes. The first two mechanisms ensure that the number of Byzantine
nodes in a node's view cannot be increased arbitrarily by attackers.
This offers protection from general Byzantine behaviors including
those resulting from botnet attacks, as defined above. The third mechanism ensures that
nodes sample their peers from a variety of address prefixes, thereby
countering \institutional attacks where the attacker controls a limited number of entire
address prefixes.

Table~\ref{table:params} shows an overview of the parameters of our
algorithm and of its environment, while Algorithm~\ref{alg:one} shows
its pseudocode.  For the sake of clarity, in the following, we use the
generic term \emph{node} to refer to protocol participants, but we use the
term \emph{peer} to refer to a node's neighbors or potential
neighbors.


\begin{table}[t]
	\scriptsize
	\begin{center}
		\caption{Parameters of the \ourmethod{} algorithm and of its environment.}
		\label{table:params}
		\begin{tabular}{l|l|l}
		  \hline
			\multicolumn{3}{c}{\textbf{Environment parameters}} \\
		  \hline
			$n$ & Number/equivalent number of nodes & 1000, 10000 \\
			$f$ & Fraction/equivalent fraction of malicious nodes & 10\%, 30\% \\
			$Q$ & Number of correct nodes & $=(1-f) n$\\
			$F$ & Attack force (described in Sec.~\ref{sec:worst-case-attack}) & $\ge 0$ \\
		  \hline
			\multicolumn{3}{c}{\textbf{Algorithm parameters}} \\
		  \hline
			$v$ & View size & 50 to 200 \\
			$\tau$ & Exchange interval & 1 time unit \\
			$\rho$ & Sampling rate (peers per time unit) & $\sim 1$ \\
			$k$ & Replacement count & up to $v/2$ \\
		  \hline
			\multicolumn{3}{c}{\textbf{Theoretical model variables}} \\
		  \hline
			$t$ & Time & \\
			$c(t)$ & Number of correct node identifiers seen & $0\le c(t)\le Q$\\
			$b(t)$ & (Equivalent) number of malicious node identifiers seen & $0\le b(t)\le fn$ \\
			$B(t)$ & Probability of sampling a Byzantine node & $=\frac{b(t)}{b(t)+c(t)}$ \\
		  \hline
		\end{tabular}
	\end{center}
	\vspace{-1em}
\end{table}

\subsection{Stubborn Chaotic Search}
\label{sec:defin-targ-graph}
\basalt nodes implicitly identify a dynamic target random graph by defining target neighbors
using a set of random ranking functions.
Then, each node greedily attempts to converge towards this implicit definition by
repeatedly exchanging neighbor lists with other peers,
discovering at each step peers that better match its ranking functions. 
In the following, we first detail the use of ranking functions to identify target neighbors. Then we discuss how nodes update these ranking functions to make the random graph dynamic. 

\begin{figure}[t]
    \centering
    \small
    \def\svgwidth{.65\columnwidth}
      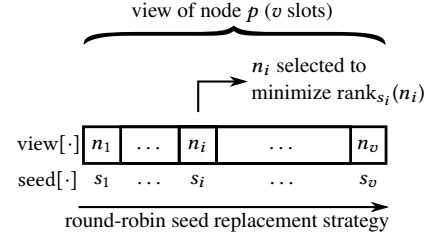

	  \vspace{.5em}
	  \caption{The mechanism of \ourmethod{} }
	  \label{fig:btrps}
	  \vspace{-1em}
  \end{figure}

\paragraph{Identifying neighbors through ranking functions}
Each node maintains a view, $\view{\cdot}$, composed of $v$
\emph{slots}. For each slot, $i \in \{1,\dots,v\}$, it chooses a
random seed, noted $\seed{i}$ (line~\ref{line:init:seed} of Algorithm~\ref{alg:one}, and
\cref{fig:btrps}) that defines a corresponding random \emph{ranking}
function, $\rank{\cdot}{\seed{i}}$. We then define a node's $i$-th
out-neighbor in the target graph as the (correct or malicious) node
$p$ that minimizes $\rank{p}{\seed{i}}$. The function $\rank{\cdot}{  \seed{i}}$ can be selected to implement specific sampling
distributions. For instance, using a simple hash function $\rank{p}{ \seed{i}}=h(\langle \seed{i}, p \rangle)$ (where angle brackets
represent a tuple) leads to a uniform sampling function, since each
peer identifier has the same probability of producing the lowest
rank. In Section~\ref{sec:sybil_biased_sampling}, we present how a
hierarchical ranking function allows \ourmethod to foil \institutional
attacks.  For simplicity, we use the shortcut of saying that a peer
$p$ better matches $\seed{i}$ than a peer $p'$ if $\rank{p}{ \seed{i}}<\rank{p'}{\seed{i}}$.

When selecting $\seed{i}$, a node cannot know the corresponding target
identifier.  Rather, it stores, in $\view{i}$, the identifier
that has so far produced the smallest value of $\rank{\view{i}}{\seed{i}}$
amongst those seen since selecting $\seed{i}$. 
At startup, each node
selects the best matching peers, $\view{i}$, from a set of
bootstrap peers (line~\ref{line:a1_bootstrap}).\footnote{We discuss the influence of the composition of this bootstrap set
in Section~\ref{sec:theory2}.}
Nodes then periodically exchange the current contents of their views
at lines~\ref{line:update:view:start}-\ref{line:update:view:end} in
order to discover new peers that can serve as better matches for the
slots in their views.  Specifically, every $\tau$ time units (exchange
interval), each correct node selects a random peer from its view and
sends it a \emph{pull} request (line~\ref{line:a1_pull}) to which the
recipient, if correct, replies by sending the contents of its current
view (line~\ref{line:a1_reply}).  Then, the node selects another peer
from its view and sends it a \emph{push} message containing its
current view (line~\ref{line:a1_push}).  When it receives the reply to
the pull request, the node greedily updates any slot $\view{i}$ that
can be brought closer to its corresponding seed, $\seed{i}$, using one
of the received identifiers
(lines~\ref{line:greedyUpdate:start}-\ref{line:greedyUpdate:end}).
The peer to which a push message was sent does the same on its side.%
\ft{We should explain in passing why we opt for a push/pull approach,
  rather than only pull or only push. I think there are papers looking
  at this: we could refer to them for detail.}

\newcommand{\algoComment}[1]{~}

\begin{algorithm}[tb]
	\small
\SetFuncSty{textsf}
\SetKwBlock{Parameters}{algorithm parameters}{end}
\SetKwBlock{Initialization}{initialization}{end}
\SetKwBlock{Function}{function}{end}
\SetKwBlock{OnReceive}{on receive}{end}
\SetKwBlock{Every}{every}{end}
\SetKwData{State}{state}
\SetKwData{Fanout}{fanout}
\SetKw{Send}{Send}
\SetKw{From}{from}
\SetKw{Sample}{Sample}
\SetKw{Or}{or}
\SetKwFunction{UpdateSample}{updateSample}
\SetKwFunction{SelectPeer}{selectPeer}
\DontPrintSemicolon
\Parameters{
	\emph{see Table~\ref{table:params}}\\
}\vspace{.7em}
\Initialization{
	\For{$i \in 1,\dots,v$}{
		$\seed{i} \gets $rand\_seed();\label{line:init:seed}
		$\view{i} \gets \bot$;
		$\counter{i} \gets 0$
	}
	$r \gets 1$;
	\UpdateSample{bootstrap\_peers}\label{line:a1_bootstrap}
}\vspace{.7em}
\Every($\tau$ time units){\label{line:update:view:start}
	$p \gets $\SelectPeer{};
	\Send $\langle$\Pull$\rangle$ \KwTo $p$ \algoComment{pull neighbor list from peer $p$}\label{line:a1_pull} \\
	$q \gets $\SelectPeer{}; 
	\Send $\langle$\Push$,\view{\cdot}\rangle$ \KwTo $q$ \algoComment{push our neighbor list to peer $q$}\label{line:a1_push}\label{line:update:view:end}
}\vspace{.7em}
\OnReceive($\langle$\Pull$\rangle\,$ \From $p$){
	\Send $\langle$\Push$,\view{\cdot}\rangle$ \KwTo $p$\label{line:a1_reply}
}
\OnReceive($\langle$\Push$,[p_1,\dots,p_v]\rangle$ \From $p$){
	\UpdateSample{$[p_1,\dots,p_v,p]$}\label{line:a1_receive}
}\vspace{.7em}
\Every($k/ \rho$ time units){\label{line:produce:sample:start}
	\For{$i = 1,\dots,k$}{
		$r \gets (r \;\mathbf{mod}\; v) + 1$ \algoComment{round-robin selection of replaced peer}\label{line:a1_select_sample}\\
		\Sample $\view{r}$ \algoComment{return it as a sample to the application}\label{line:a1_sample}\\
		$\seed{r} \gets $rand\_seed() \algoComment{select new seed in order to sample another peer}\label{line:a1_reset}\label{line:produce:sample:end}\\
	}
	\UpdateSample{$\view{\cdot}$}\algoComment{select best matching peers from current neighbors}\label{line:a1_reset2}
}\vspace{.7em}
\Function(\UpdateSample{$[p_1,\dots,p_m]$}){
	\For{$i \in 1,\dots,v$, $p \in [p_1,\dots,p_m]$}{
			\If{$p = \view{i}$}{
				$\counter{i} \gets \counter{i} + 1$	\algoComment{if $p$ was already best matching peer, increase hit counter}\label{line:a1_inc_ci}
			}
			\ElseIf{$\view{i} = \bot$ \Or $\rank{p}{\seed{i}} < \rank{\view{i}}{\seed{i}}$\label{line:greedyUpdate:start}}{
				\label{line:a1_replace_ni}
				$\view{i} \gets p$; $\counter{i} \gets 1$\label{line:a1_replace_ni2}\label{line:greedyUpdate:end}\label{line:a1_reset_ci}
			}
	}
}
\Function(\SelectPeer{}){
	$i \in \mathrm{argmin}_{j=1}^v (\counter{j})$ \algoComment{select any peer with the minimal value of $\counter{i}$}\label{line:a1_select_min_ci}\\
	$\counter{i} \gets \counter{i} + 1$ \algoComment{increase its hit counter}\label{line:a1_inc_ci_selected}\\
	\Return \view{i}
}
\caption{The \ourmethod{} algorithm}
\label{alg:one}
\end{algorithm}

\paragraph{Making the graph dynamic}

To generate a dynamic random graph and enable nodes to continuously
generate fresh samples from the network, nodes regularly reset some of their
seeds to new random values.  
This periodic operation,
at lines~\ref{line:produce:sample:start}-\ref{line:a1_reset2},
first provides the application with $k$ peer identifiers representing
a random sample of the network (line~\ref{line:a1_sample}),
and then resets the $k$ corresponding slots by selecting
new seeds $\seed{r}$ (line~\ref{line:a1_reset}).
These $k$ slots are
selected in a round-robin fashion every $k/\rho$ time units. This
yields $\rho$ random samples per time unit on average as indicated in
Table~\ref{table:params}. 
It then sets the
corresponding entries, $\view{r}$, to the identifiers from the current
view that best match the new seeds (line~\ref{line:a1_reset2}).  When
the algorithm returns $\view{r}$ as a sample to the application,
$\view{r}$ effectively results from a random selection amongst all the
peer identifiers received since the last reset of \seed{r}.  A node
has no way of knowing if it has found the peer $p$ that best matches
$\seed{r}$ globally (i.e.~its target neighbor in the random graph),
but selecting which seeds to reset in a round-robin fashion and by sampling
$\view{r}$ just before resetting $\seed{r}$, the algorithm ensures
that a maximum number of identifiers have been seen for each seed when
returning the corresponding sample. This optimizes the randomness of
the sample for a given budget of peer exchanges and view
size.%
\footnote{Other than using more memory, increasing the view size
  has a non-negligible networking cost as one would typically keep an
  open TCP connection ready for each peer of the current view.
  Alternatively, the samples' randomness could be increased
  by keeping a log of recently closed connections and re-injecting
  these peer identifiers when selecting new seeds.}

Parameter $\rho$ controls the number of random samples per time unit,
and so the number of slots whose seeds are refreshed at each time
unit. With a view size of $v$, this means that each slot is refreshed
on average every $v/\rho$ time units. The value of $v/\rho$ must
therefore be large enough with respect to the exchange interval,
$\tau$.  Parameter $k$ controls, instead, the number of slots that are
reset at the same time.  A large value of $k$ causes the algorithm to
explore many slots in parallel, thereby obtaining more diverse samples
that help the $k$ slots converge faster together. A small value of $k$
(e.g. $k=1$), instead, causes the exploration to occur with most slots
in a quasi converged state.  This increases the probability of
contacting peers that have already been contacted recently, thereby
leading to slower convergence for the $k$ unconverged slots. Our
experiments by simulation confirm that \ourmethod better resists the
presence of Byzantine peers using a batch sampling-and-replacement
strategy where $k$ can be as high as $v/2$ (in which case $k/\rho$,
like $v/\rho$, must also be at least several exchange intervals,
$\tau$), rather than replacing seeds one by one (i.e.~setting
$k=1$).\ft{Any downside to a large $k$? If not, why not have $k=v$?}

\subsection{Hit Counter Hardening Mechanism}\label{sec:hit_counter}
\label{sec:hard-mech-using}
The graph-generation mechanism described above prevents Byzantine peers from influencing the target graph,
and thus the views of correct nodes once the network has converged.
However, depending on the speeds at which correct nodes discover  other correct or Byzantine peers, 
their intermediate views may suffer from a bias in favor of Byzantine
nodes.  If this happens, the algorithm will tend to select malicious
nodes to push to and pull from (lines~\ref{line:a1_pull}
and~\ref{line:a1_push}), further slowing down convergence.

\ourmethod mitigates this issue by introducing a hit-counter mechanism that effectively makes the protocol harder to attack. 
Each node maintains a \emph{hit counter} variable, $\counter{i}$, for each  slot $i \in \{1,\dots,v\}$ in its view. A node sets 
$\counter{i}$  to $1$ when initializing the slot, as well as whenever updating $\view{i}$ with a peer that better matches the corresponding seed (line~\ref{line:a1_reset_ci}).
Every time a node receives another peer's view that also contains peer $\view{i}$,
it increases  $\counter{i}$  by one (line~\ref{line:a1_inc_ci}).

When deciding which neighbors to contact,
a node always selects one of the peers with the lowest value of $\counter{i}$ (line~\ref{line:a1_select_min_ci}).
Finally, the node increases the hit counter of the selected peer by $1$ to make it less likely to be selected  the next time
(line~\ref{line:a1_inc_ci_selected}).

This mechanism has no impact on honest nodes as they should each appear as often in expectation.
However, it creates a trade-off for (possibly colluding) malicious nodes that try to  be over-represented,
as nodes attempting to appear more often will automatically be contacted less.
We further discuss this aspect and the possibility of attacks on the hit counter mechanism in Section~\ref{sec:analys-hard-mech}.

\subsection{Hierarchical ranking}
\ft{Alternative title: `Hierarchical sampling against \institutional attacks'}%
\label{sec:sybil_biased_sampling}

\ft{In the following I've tried to link the discussion to the algorithm we have just presented, and to focus specifically on \institutional attacks, which biased sampling targets. This raises the question of where we should discuss botnet attacks. Maybe earlier, while presenting the algorithm? Or here too, to explain that a non-uniform sampling does not reduce the robustness against botnet attacks, which we assume are follow the same distribution as honest peers. (Do we explain this somewhere. If not, should we?)}
Central to Algorithm~\ref{alg:one}, the function $\rank{\cdot}{\seed{i}}$ induces a specific sampling distribution of node identifiers.
\alt{Using Algorithm~\ref{alg:one}, honest nodes seek to converge towards a graph defined by the function $\rank{\cdot}{\seed{i}}$, which induces a specific sampling distribution of node identifiers.}%
For instance, using simply a hashing function for $\rank{\cdot}{\seed{i}}$ yields uniform node sampling. Unfortunately, 
uniform 
sampling
makes it relatively easy for \institutional attackers 
to
gain an overwhelming 
influence in the system, by taking the control of large IP address blocks 
to implement Sybil attacks (Section~\ref{sec:sybil_attack_def}).
For instance, as we will see in Section~\ref{sec:numer-analys-case}, controlling one of the largest ISPs\ft{Is the largest AS indeed an ISP?} would grant an attacker about $10^8$ IPv4 addresses, a number large enough to thwart most existing decentralized BFT systems.
\ft{Not sure we need to provide concrete details here.}%
\ft{Would be nice to inject a bit \institutional attacks here, as this is way really ranking functions target rather than botnet attacks.}%
Such attacks, are, however, heavily concentrated in a limited number of address ranges by design ($\sim$5700 address blocks in the above example).
The key idea for countering
them therefore consists in not sampling peers uniformly, but
using ranking functions 
that induce\ft{inject?} some diversity in the sampled node identifiers, and thus reduce the probability of sampling
Sybil peers. 

\paragraph{Ranking functions and target distributions}
Formally, let $S$ be a uniform random variable on 256-bit integers, which
corresponds to the sampling of a seed.
Let $X$ be the random variable corresponding to the \emph{best
matching peer sample} for $S$, defined as:
\begin{dmath}X = \mathrm{argmin}_{p \in \mathcal{N}} \rank{p}{S}\label{eq:def:attack:power}\end{dmath}
where $\mathcal{N}$ denotes the set of all network nodes.  Depending
on the definition of $\rank{p}{S}$, $X$ can implement a
specific probability distribution on network nodes. This allows us to
define the attacker's \emph{power}, $f$, as the probability of $X$ being a
malicious node given a specific ranking function $\rank{\cdot}{S}$. 
\df{We probably want to remove the following or edit}
If $\rank{\cdot}{S}$ is a simple hashing function,
$\rank{p}{S}=h(\langle S, p \rangle)$, nodes are selected uniformly, and
$f$ corresponds to the fraction $\varphi$
of malicious nodes in the system.
In the more general case, this probability is no longer equal to $\varphi$
but, as we will see in Section~\ref{sec:intro_analysis}, it plays the same role in our analysis,
thus we will also call $f$ an \emph{equivalent fraction} of malicious nodes.

\paragraph{Selecting the ranking function}
In order to counter \institutional Sybil attacks, we need to select a ranking
function that minimizes $f$ by giving malicious nodes a low probability of being
selected as best-matching peers (i.e. chosen by the distribution $X$).
\ourmethod adopts a ranking function that spreads sampled peers amongst different subnets by exploiting the structure of IP addresses. IP addresses can indeed usually be decomposed in two parts, a prefix, that designates the \emph{subnet} to which the
address belongs (linked to a given Internet service
provider), and a local part that identifies a node within that
subnet.


\ft{I've removed this bit about assumptions about what is easy or not, as not sure this works well now that we've defined our two attack models (\institutional and botnet).
`Our method is based on the assumption that it is easier for an attacker
to gain control of a given subnet than of a large set of addresses spread
amongst many different subnets.
This assumption is realized in our first attack scenario,
an \institutional attack
where an institution makes use of all the IP addresses it owns to attack
the peer sampling algorithm.
As we will see below, our method greatly reduces the attacker's
power in this scenario.
This assumption is however not necessarily realized in the case of a botnet attack,
where infected machines can be located in many different subnets.
However botnets usually contain less nodes, as we will see below,
and their scope can therefore be reduced simply by having as many honest nodes
as possible in the network.'
}

\paragraph{A first grouped ranking function}
To illustrate this intuition,
suppose that $G(p)$ corresponds to 
the prefix of a given length of the IP address,
or a country code determined from the address.
The following ranking function (based on a lexicographical ordering on values)
can be used to sample uniformly amongst the different values of property $G(p)$,
and then uniformly amongst all the peers that have the selected value of $G(p)$:
  $$\rank{p}{S} = \big\langle\, h(\langle S, G(p) \rangle)\,,\,h(\langle S, p\rangle) \,\big\rangle$$

  Using such a ranking function makes an attack against \ourmethod{}
  harder. In order to gain a power of $f$ in the network, a malicious
  entity would need to control a large number of nodes at least in a
  fraction $f$ of all the values of $G(p)$ where network nodes exist.
  For instance, consider an attacker that owns a full IP address
  block.  
  Uniform node sampling gives the attacker a power of
  $f=\frac{q}{n}$, where $q$ is the size of the IP block and $n$ the
  total number of nodes in the network.  In the group-based sampling
  model, since an address block is usually associated with a single
  group (a single country, a single IP address prefix), the attacker
  only has a power of $f=\frac{1}{|G|}\frac{q}{g}$, where $|G|$ is the
  number of different groups, and $g$ is the number of nodes present
  in the particular group of the attacker's address block.  This
  attacking power is trivially bounded by $\frac{1}{|G|}$, and the
  only way to increase it consists in taking control of many IP addresses
  in other groups, making such an attack much more costly.

\paragraph{\ourmethod's hierarchical ranking function}
In \ourmethod, we take the grouping approach described above one step further. We adopt a hierarchical  ranking function that descends the address hierarchy
by sampling uniformly at levels /8, then /16, then /24, and then finally at the level of individual addresses, defined as follows:
\begin{dmath}\label{eq:rank:def}
\rank{p}{S} = \Big\langle h\big(\langle S, p_0^8 \rangle\big), h\big(\langle S, p_0^{16}\rangle\big),
             h\big(\langle S, p_0^{24} \rangle\big), h\big(\langle S, p\rangle\big) \Big\rangle
\end{dmath}%
where $p_0^i$ corresponds to the prefix constituted of the $i$ most
significant bits of $p$'s IP address.
The efficiency of this ranking function in countering \institutional Sybil attacks
is demonstrated numerically in Section~\ref{sec:numer-analys-case}.

\section{Theoretical Analysis}%
\label{sec:theory}%

We now use a theoretical continuous model to estimate
the value of $B(t)$,
the probability at a time $t$ that a given slot of a correct process contains a Byzantine peer identifier,
as a function of
$f$, the attacker's power.\ft{I'd single out $f$ to link this section with the discussion in Sec 3. The fact that protocol parameters and scenario parameters play a role too is obvious, so we can drop them I think.}
The use of $f$ allows us to apply the same analysis to \infrastructure and botnet attacks in a unified reasoning.%
\ft{Removing this bit, as now the paper is no longer organized around the permissioned and permissionless cases, but around \infrastructure and botnet attacks. `The use of a sampling distribution defined by a ranking function
allows us to bring back the case of a permissionless network to
that a permissioned scenario with a known fraction of malicious nodes.
We will thus base our analysis on the general principle of a permissioned setting.'}

\subsection{Parameters, Notations and Assumptions}
\label{sec:intro_analysis}

\subsubsection{Scenario parameters and node distribution}
\label{sec:scenario_parameters}

\ft{I had written the following, but I don't think we need it: `In all attack scenarios we assume that honest nodes are uniformly distributed across the IP address space, and in particular that the proportion of honest nodes in a given prefix is proportional to this prefix's block size.'}

We first consider an ideal `uniform' botnet attack, in which Byzantine identifiers follow the same distribution as those of honest nodes. This situation corresponds for instance to a scenario in which a botnet indiscriminately targets the same kind of nodes (e.g. personal machines) as those making up the rest of the system.
In this case, the attacker's power $f$
that we introduced in Section~\ref{sec:sybil_biased_sampling} is simply equal to the fraction of Byzantine nodes in the network, and is independent of \ourmethod's hierarchical ranking function.  
To analyze this attack, we note $n$ the total number
of network nodes (i.e. the network size), 
the product
$fn$ denotes the number of Byzantine nodes, and $Q=(1-f)n$ denotes the
number of correct nodes.

In the case of an \infrastructure attack, the attacker's power $f$ depends on the distribution of the address blocks it controls,
and
represents the probability of
selecting a Byzantine identifier using the ranking function
$\rank{p}{S}$ in the hypothetical case that all identifiers in
the network are known (Eq.~\ref{eq:def:attack:power}). In this scenario, we define $n$ as an \emph{equivalent network size},
defined as $n=\frac{Q}{1-f}$, where $Q$ still denotes the number of correct nodes.
These definitions satisfy the equality $Q=(1-f)n$, as in the
(uniform) botnet attack, and will allow us to apply the same analysis seamlessly.

\newcommand{\nbBlock}{\ensuremath{m}}
\newcommand{\nbBlockOn}[1]{\ensuremath{\nbBlock_{#1}}}
\newcommand{\nbBlockRemain}{\nbBlockOn{C}^\star}


The two above scenarios represents the extreme cases of a wider spectrum of attacks. In particular, botnet attack might not be uniform, as the identifiers controlled by a botnet might be biased towards certain blocks (e.g. in the case of botnet built by targeting certain organization, or specific vulnerabilities) that differ from those of honest nodes. In such hybrid cases, the reasoning for \infrastructure attacks applies.

\subsubsection{Notations}

\newcommand{\correctSet}{\mathcal{C}}
\newcommand{\byzSet}{\mathcal{B}}
\newcommand{\myequationSize}{\fontsize{9pt}{\baselineskip}}

The probability $B(t)$ of selecting a Byzantine node in a given slot 
of a node $p$ at time $t$
depends on two sets of identifiers:
the set of correct identifiers seen at a time $t$ by $p$ on this slot since the last reset, noted $\correctSet(t)$,
and the set of Byzantine identifiers seen by $p$ over the same period, noted $\byzSet(t)$.

One key observation is that, for a fixed $\correctSet(t)$, $B(t)$ increases as $p$ hears of new Byzantine identifiers and $\byzSet(t)$ grows, 
i.e. $\correctSet(t)=\correctSet(t')\wedge \byzSet(t)\subseteq\byzSet(t') \implies B(t)\le B(t')$,
so that for a given $\correctSet(t)$, $B(t)$ is maximum when the node $p$ has learned all Byzantine identifiers  circulating in the system.


In the following analysis, we therefore assume a worst case scenario in which correct nodes have been flooded with all existing Byzantine identifiers. (We discuss the actual implementation of this worst case scenario in Section~\ref{sec:worst-case-attack}.)
For botnet attacks, we have assumed that correct and Byzantine identifiers follow the same distribution, implying that 
\newcommand{\bmax}{b_{\max}}
\begin{equation}\myequationSize\label{eq:C:from:bmax}
C(t)=\frac{c(t)}{\bmax+c(t)},
\end{equation}
where $\bmax$ is the total number of Byzantine identifiers, i.e.~$\bmax=fn$. (See Appendix~\ref{sec:botnet-attack} for a detailed derivation.)

For \institutional attacks, we assume that the distribution of correct nodes
is independent of the sampling distribution introduced by $\mathrm{rank}()$,
and we approximate $C(t)$
using the same form as Eq.~\ref{eq:C:from:bmax},
where $\bmax$ becomes an \emph{equivalent} number of Byzantine identifiers.
Considering the case when $p$ knows all correct nodes ($c(t)=Q$),
and having defined $n=\frac{Q}{1-f}$, 
we derive $\bmax=f\times \frac{Q}{1-f}$, and hence $\bmax=fn$ in this case as well,
where $n$ is now the equivalent network size introduced above.

In both attacks, the probability of selecting a Byzantine node,
$B(t)$, becomes therefore driven by the number of correct identifiers known to $p$, $c(t)=|\correctSet(t)|$,
and the same system equation can be used to study both cases (modulo the redefinition of $n$ for \institutional attacks).

\subsubsection{Assumptions} For simplicity, we study a version of \ourmethod without the
hit counter-based hardening mechanism, and later discuss its impact in
Section~\ref{sec:analys-hard-mech}. Algorithm ~\ref{alg:two} shows the
pseudocode corresponding to the hit counter-less version being
analyzed. 
To approximate the system's behavior, we will reason using the mean
values of $c(t)$ 
over all nodes and slots, and assume that
the values of individual nodes tend to concentrate around their means
in practice with high probability, as is usually the case in such
stochastic systems.

\subsection{Analysis of the Core Mechanism}
\label{sec:theory_alg_one}
We first 
discuss in more detail the worst-case attack on the \ourmethod{} algorithm.
We then study the risk of a node becoming isolated under this attack model (i.e.~of an Eclipse attack succeeding),
before moving on to studying the convergence properties of \ourmethod{}
assuming no node is ever isolated.

\subsubsection{Identifying the Worst-Case Attack}
\label{sec:worst-case-attack}

To identify the worst case attack, we observe that attackers cannot influence the choices correct nodes make (at line~\ref{line:a2_select_peer}
of Algorithm~\ref{alg:two}, and at lines~\ref{line:a1_select_sample}-\ref{line:produce:sample:end} of Algorithm~\ref{alg:one});
thus they can only manipulate the peer-sampling
process by increasing their representation in the views of correct
nodes, i.e. the value of $B(t)$. The fact that  $B(t)$ grows with the
set of Byzantine identifiers the node is aware of, $\byzSet(t)$, 
suggests that the
worst case scenario arises when Byzantine nodes flood the network
with their identifiers in order to increase $\byzSet(t)$ as much as possible.
We model this attack scenario as follows: 

\begin{itemize}
\item A malicious node that receives a pull request returns a view
  composed of $v$ nodes selected uniformly at random amongst the
  malicious nodes. 
\item Regularly, a malicious node sends a push request to randomly
  selected correct peers, containing similarly a view of $v$ uniformly
  random malicious peers.
\end{itemize}

\begin{algorithm}[t]
	\small
\SetFuncSty{textsf}
\SetKwBlock{Initialization}{initialization}{end}
\SetKwBlock{Function}{function}{end}
\SetKwBlock{OnReceive}{on receive}{end}
\SetKwBlock{Every}{every}{end}
\SetKwData{State}{state}
\SetKwData{Fanout}{fanout}
\SetKw{Send}{Send}
\SetKw{From}{from}
\SetKw{Sample}{Sample}
\SetKw{Or}{or}
\SetKwFunction{UpdateSample}{updateSample}
\SetKwFunction{SelectPeer}{selectPeer}
\DontPrintSemicolon
\Function(\UpdateSample{$[p_1,\dots,p_v]$}){
	\For{$i \in 1,\dots,v$}{
		\For{$p \in [p_1,\dots,p_v]$}{
			\lIf{$\view{i} = \bot$ \Or $h(\langle \seed{i}, p \rangle) < h(\langle \seed{i}, \view{i} \rangle)$}{\label{line:a2_replace_ni}
				$\view{i} \gets p$\label{line:a2_replace_ni2}
			}
		}
	}
}
\Function(\SelectPeer{}){
	$i \gets $rand$(1,\dots,v)$;\label{line:a2_select_peer}
	\Return \view{i}
}
	\caption{Simplification of Algorithm~\ref{alg:one} for theoretical analysis of Section~\ref{sec:theory_alg_one}}
\label{alg:two}
\end{algorithm}

 We define the \emph{force} of the attack, $F$ (distinct from the attacker's power, $f$), as the ratio between
 the number of push requests sent by a Byzantine node and number of
 push requests sent by a correct node in a given time interval. For
 example, if a Byzantine node sends push requests at the same rate as
 correct nodes, a force of $F$ corresponds to sending requests to $F$
 distinct correct nodes rather than to only one. Alternatively, the
 force of the attack can also model a situation in which Byzantine
 nodes send requests more often, or where the network loses more
 messages from correct nodes than from Byzantine ones.

 The worst case corresponds to an arbitrarily large value of $F$,
 arising when correct nodes receive all the identifiers of Byzantine
 peers in any arbitrarily small (but non-empty) time interval.
 This means that apart from the initial state,
 $\byzSet(t)$ is constant, and
 its effect can be captured by the term $\bmax=fn$ to compute the probabilities of selecting a correct (resp. Byzantine) peer in a slot of a node's view.
 We recall
 that $fn$ represents the total/equivalent number of Byzantine nodes
 depending on the attack considered. The analysis that follows shows that even
 in this case, \ourmethod{} causes $B$ to converge
 to a value that is only slightly larger than the attacker's power, $f$.
 The experimental results of Section~\ref{sec:exp_simulation} analyze instead
 the actual performance with finite values of $F$.

\subsubsection{Bounding the Probability of Isolation}
\label{sec:theory2}
We start by showing that nodes have a low probability of being
isolated. Isolation can happen in two ways: either when a node joins
the network for the first time, or when it evicts all correct peers
from its view and replaces them with Byzantine peers.

\paragraph{Isolated joining node}

In the first case, the unfortunate joining node receives all of the identifiers of Byzantine nodes as soon as it joins. At time $\epsilon$ after joining we have $\byzSet(t)$ becomes maximal 
and $c(\epsilon)=(1-f_0)I$, where $f_0$ is the
fraction of Byzantine nodes in the bootstrap sample and $I$ is the size of the bootstrap sample.
Since we defined $B(t)$ as the probability of a given slot
in the view being occupied a Byzantine peer, we can write the
probability that a node has only Byzantine neighbors as $B(t)^v$.

{\myequationSize
\begin{equation}
B(t)^v = \left(\frac{\bmax}{\bmax+c}\right)^v = \left(\frac{1}{1 + \left(1-f_0\right)\frac{I}{fn}}\right)^v
\end{equation}
}

We can reduce this probability exponentially by increasing $v$, by
increasing $I$ or by assuming a lower $f_0$.  For instance, supposing
$f_0=50\%$ of malicious nodes in our bootstrap peer list, by taking a
view size of $v=200$ and a bootstrap peer list size $25\%$ of the
number of malicious nodes in the network ($I=\frac{1}{4}f n$), this
probability becomes smaller than $10^{-10}$.
Supposing for instance a network of size $n=10000$ with a fraction $f=0.1$
of Byzantine nodes, this only requires a bootstrap set of size $I=250$ nodes,
of which only $125$ are required to be correct.

\paragraph{Convergence to isolated state}
\label{sec:convergence-isol-state}
The second way for a node to become isolated results from resetting
the seeds for the slots that still contain correct peers to new seeds
that select Byzantine nodes.  When such a reset occurs, the
probability that all of the non-reset slots are already owned by
Byzantine peers is equal to $B(t)^{v-k}=\left(\frac{\bmax}{\bmax+c(t)}\right)^{v-k}$.
When the number of correct nodes seen locally, $c(t)$, is large enough,
this probability is negligible.

Let us now study the value of $c(t)$ at the time of a reset, depending
on the value of $c(t)$ at the time of the previous reset.  For this
analysis, we look at a single node of the network and make the
hypothesis that other network nodes are well-converged. As we discuss
in Sections~\ref{sec:theory3} and~\ref{sec:exp_simulation}, this implies
that the fraction of Byzantine nodes in their views approaches $f$ with
appropriate algorithm parameters.
We write $c_0$ the value of $c(t)$ at the previous reset.
The expected number of correct peer identifiers received during the period between
the two resets is lower bounded by $\frac{k}{\rho} \frac{v}{\tau} \frac{c_0}{fn+c_0} (1-f)$.
If we write $\Delta c$ the corresponding increase in $c(t)$, i.e.~the number
of \emph{distinct} correct peer identifiers received during this time period,
we obtain:
{\fontsize{9pt}{\baselineskip}
\begin{equation}
	\Delta c \ge \frac{k v c_0 (1-f) (Q - c_0)}{Q \tau \rho (fn+c_0) + k v c_0 (1-f)}
	\label{eqn:convergence-isol-state-result}
\end{equation}
}
\noindent (see Appendix~\ref{sec:apx-convergence-isol-state} for the full derivation).

Suppose for instance a network of $n=10000$ nodes with a proportion $f=0.1$
of malicious nodes, with algorithm parameters $v=100$ and $k=50$.
In this system, taking $\tau=1$ and $\rho=1$,
and supposing that the node we are considering has just joined the network and
knows only of $c_0=f_0 \frac{1}{4} f n = 125$ correct node identifiers,
we obtain that $\Delta c \ge 467$, i.e.~$c(t)$ at the next reset is expected to be at least $592$.
$B(t)^{v-k}$ is smaller than $10^{-10}$ as soon as
the number $c(t)$ of correct node identifiers seen is more than $585$.
In other words, the probability that the node becomes isolated at the next reset
is negligible.
This guarantee can be made even stronger by increasing the view size $v$.
Moreover, if the node is already in a better-converged state with a relatively large $c_0$,
the probability of becoming isolated during a reset becomes even smaller.

\subsubsection{Non-Isolated Execution}
\label{sec:theory3}
Now that we have shown that the probability of a node becoming isolated can be made arbitrarily low, we make the following assumption in the rest of the analysis. 
\begin{assumption}
No node is isolated, and all nodes have at least some correct neighbors. In particular,  $B(t)^v$ is negligible at all times $t$. 
\label{ass:non-isolated}
\end{assumption}

\paragraph{Deriving a continuous model}

When Assumption~\ref{ass:non-isolated} holds, and in the worst-case
scenario discussed above (Byzantine nodes have propagated all their
identities to all correct nodes) $\byzSet(t)$ is constant, and its effect captured by the term $\bmax$, which allows
us to write the evolution of $c(t)$ over time as a differential
equation, as the sum of contributions resulting from the various
parts of the system.

\begin{itemize}
\item \emph{Pull exchange}: every $\tau$ rounds, a node pulls from one
  peer in its view, which replies by sending $v$ node identifiers.
  With probability $B(t)$, the node contacts a Byzantine peer. In this
  case, it receives only Byzantine peer identifiers that it is already
  aware of (by the worst-case assumption $\bmax=fn$).  With probability
  $C(t)$, the node contacts instead a correct peer. In this case, each
  returned identifier will itself be correct with probability $C(t)$,
  and if correct, it will have a probability $\frac{c(t)}{(1-f)n}$ of
  being already known ($(1-f)n$ being the total number of correct
  nodes). Thus we can express the variation of $c(t)$ over time as a
  result of a pull operation as follows:
                {\myequationSize
                  \begin{dmath*}
			\frac{dc}{dt}=\frac{1}{\tau}\left[C(t)^2 v \left(1-\frac{c(t)}{(1-f)n}\right) \right].
		\end{dmath*}
		}
	
              \item \emph{Push exchange}: every $\tau$ rounds, a node
                pushes to a random node in its view.  This push has a
                $C(t)$ probability of being sent to a correct node. In
                this case we can apply the same reasoning as above and
                derive the same contribution to $\frac{dc}{dt}$.

              \item \emph{Sampling and view renewal}: every $\rho$
                rounds, a node resets one of its $v$ slots and forgets
                the identifiers collected for this slot. Let us write
                $c(t)$ as $c(t)=\frac{1}{v}\sum_{i=1}^{v}c_i(t)$,
              where $c_i(t)$~ is the number of correct nodes taken
              into account by slot $i$. Then a single $c_{i}$ is set to
              zero every $\rho$ rounds. On average, this yields the
              following contribution to $\frac{dc}{dt}$:
		{\myequationSize
			\begin{dmath*}
				\frac{dc}{dt} = -\rho \frac{c(t)}{v}.
			\end{dmath*}
		}

\end{itemize}
\noindent
By summing all three above contributions, we obtain our final  differential equation:
{\myequationSize
\begin{dmath}
		\frac{dc}{dt} = \frac{1}{\tau}\left[2 C(t)^2 v \left(1-\frac{c(t)}{(1-f)n}\right) \right] - \rho\frac{c(t)}{v}.
		\label{eqn:dcdt}
	\end{dmath}
}

\paragraph{Solving the continuous model}
We now solve Equation~\ref{eqn:dcdt} under
Assumption~\ref{ass:non-isolated} and show that the network converges to a state
where the proportion $B$ of Byzantine peers in nodes' views is small
even for arbitrarily large values of the attack force, $F$.
To this end, we can express  $\frac{dB}{dt}$ as $\frac{dB}{dt}=-\frac{\bmax}{(\bmax+c)^2}\frac{dc}{dt}$, and by substituting $\frac{dc}{dt}$ from Equation~\ref{eqn:dcdt}, we obtain:
{\myequationSize
\begin{dmath}
	\frac{dB}{dt}
	= B(1-B)\left(\frac{\rho}{v}-\frac{2v(1-B)(B-f)}{\tau f (1-f) n}\right)
	\label{eqn:dbdt2}
\end{dmath}
}

To study the constant regime of this system, we write
$\frac{dB}{dt}=0$ and exclude the solutions $B=0$, which is not
compatible with $\bmax=fn$, and $B=1$, which corresponds to the case
where Byzantine nodes take over the whole network.  We also simplify by
setting $\tau=1$ as its role is symmetrical with that of $\rho$.  We
obtain after a few steps:
{\myequationSize
\begin{dmath}
	(1-B)(B-f)=\frac{\rho f (1-f)n}{2v^2}.
\label{eq:solution}
\end{dmath}
}

The equation exhibits two roots $B_1 < B_2$.
{\myequationSize
  \begin{dmath}
	  B_{1,2}=\frac{1}{2}\left(1+f\mp\sqrt{(1-f)^2-2\frac{\rho f (1-f) n}{v^2}}\right)
    \label{eq:solved}
  \end{dmath}
}

\noindent 
When the quantity on the right-hand side of Equation~\ref{eq:solution}
approaches zero, $B_1$ approaches $f$ from above, while $B_2$
approaches $1$ from below. Since $\frac{dB}{dt}>0$ for $B<B_1$ and
$B>B_2$, while $\frac{dB}{dt}<0$ for $B_1<B<B_2$, $B_1$ corresponds to
a stable equilibrium, while $B_2$ corresponds to an unstable
one. So we focus our analysis on $B_1$.

With respect to $B_1$, the right-hand side of
Equation~\ref{eq:solution} represents the difference between the
proportion of malicious peers in nodes' views, $B(t)$, and their
overall proportion in the network, $f$. Ideally, we want to keep this
quantity as small as possible, making $B$ only slightly larger than
$f$.

To this end, we observe that the term $\frac{\rho f (1-f) n}{2v^2}$ shrinks
proportionally to the square of the view size, $v^2$.  Thus, choosing
a large enough view size allows the network to converge to a globally
well mixed state where Byzantine nodes control only slightly more
peers in the view than their overall proportion in the network.
Moreover,
in order to obtain the same stable state value of $B$, 
for fixed values of $f$ and $n$, 
the view size $v$
should grow proportionally to the square root of the sampling rate
$\sqrt{\rho}$, while, for fixed values of $f$ and $\rho$, it
needs to increase proportionally to $\sqrt{n}$.

\subsection{Analysis of the Hardening Mechanism}
\label{sec:analys-hard-mech}
\ourmethod's hit counter-based hardening mechanism allows nodes to
detect which peers have appeared more often in incoming messages, and
prioritize other peers for network exploration.  In the case of a
standard attack, where malicious peers flood their own identifiers, the
hit counter favors the choice of correct peers over malicious ones.

However, the fact that we analyzed a simplified version of \ourmethod
without the hardening mechanism raises the legitimate question of
whether the hit counter may degrade the security of the approach by enabling
some other attack.  To answer this question, let us consider a
malicious node or a coalition of malicious nodes that want to influence
the sampling operations performed by a correct node.

We start by observing that malicious nodes can neither write nor read
the local memories of correct nodes.  So they cannot influence
sampling operations directly: their only strategy consists in trying
to convince a target correct node that some of the correct peers in
its view are malicious by increasing their hit counters. To this end,
malicious nodes can repeatedly advertise the identifiers of correct
peers. But again, they cannot guess which correct peers are in the
target correct node's view slots. So their only option consists in
advertising a possibly large random set of correct peers in the hope
that some of them will be in the correct node's view.

But even this turns out to be counterproductive. If malicious nodes
advertise a large number of correct identifiers, it is indeed possible
that they may increase the hit counter of some entries in the target's
view. But they do so at the cost of increasing their target's value of
$c(t)$, i.e. the number of correct peers known to it. Since we already
assumed the worst case scenario of $\bmax=fn$, the increase in $c(t)$
can only decrease $B(t)$ thereby making the attack counterproductive.

\begin{table}[t]
	\small
	\begin{center}
		\caption{\Institutional attack: power $f$ of the adversary, defined as the equivalent fraction of malicious nodes, for different sampling methods,
		supposing biggest Internet service provider as an adversary ($\sim 10^8$ IP addresses, distributed over 5739 blocks), on the IPv4 network.
		Data sourced from the GeoLite2 Block/ASN dataset.}
		\label{table:equiv-f-sampling-method}
    \begin{tabular}{|l|c|c|c|}
		  \hline
		\textbf{\# of honest IPs ($Q$)} & \textbf{100} & \textbf{1000} & \textbf{10000} \\
			\hline
        Uniform & 99.9999\% & 99.999\% & 99.99\% \\
      \hline
        By /8 prefix & 49\% & 28\% & 27\% \\
      \hline
        By /16 prefix & 95\% & 64\% & 17\% \\
      \hline
        By /24 prefix & 99.98\% & 99.8\% & 98\% \\
      \hline
        Hierarchical & \textbf{47\%} & \textbf{21\%} & \textbf{10\%} \\
      \hline
		\end{tabular}
	\end{center}
	\vspace{-1em}
\end{table}

\begin{figure}[t]
	\centering
	\includegraphics[width=.35\textwidth]{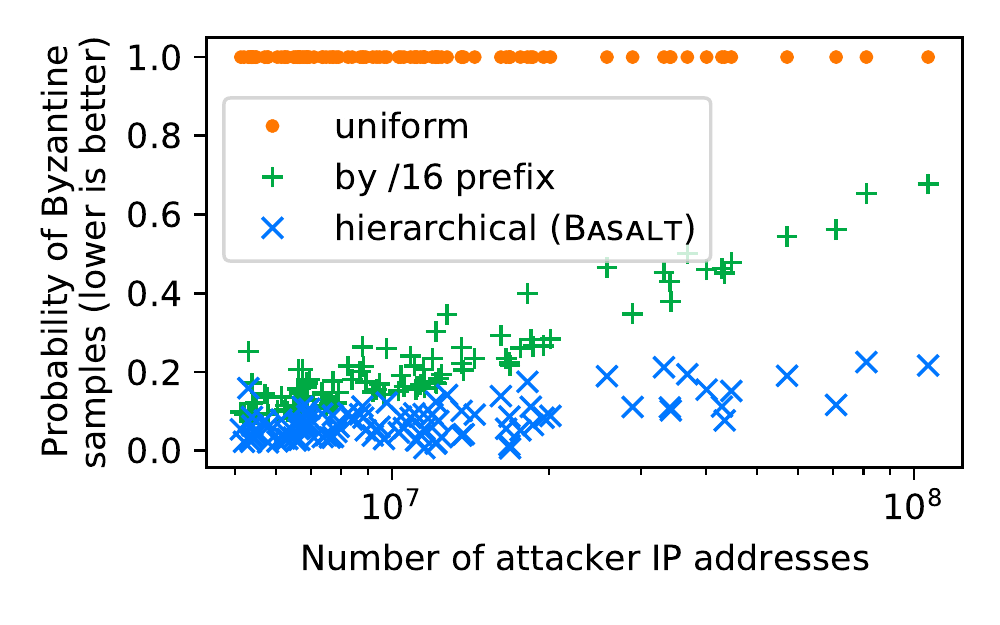}
	\caption{\Institutional attack:
		probability of sampling a Byzantine node with different ranking functions,
		calculated for the 100 biggest Internet ASes supposing that
		each of them is the attacker, and that 1000 honest nodes are
		uniformly spread amongst remaining IP address space.
		Data sourced from the GeoLite2 Block/ASN dataset.
		Horizontal axis: number of IP addresses controlled by the AS.
		The sampling probability is calculated from the equivalent fraction
		by using Equation~\ref{eq:solved},
		with a view size of $v=100$ and a sampling rate of $\rho=1$.}
	\label{fig:geolite_AS_B_n1000}
	\vspace{-1em}
\end{figure}

\subsection{Numerical Analysis: \Institutional Attacks}
\label{sec:numer-analys-case}

To illustrate the robustness of 
\ourmethod's hierarchical ranking function
against an \institutional attack,
we calculate the power $f$ (here an equivalent fraction of malicious nodes)
of a real-world attacker using data from the GeoLite2 Block/ASN dataset~\cite{geolite_asn_url},
and use the equilibrium formula of Equation~\ref{eq:solved} ($B_1$) to compute the proportion of Byzantine nodes that \ourmethod would return in such a scenario.
We assume that the attacker is an internet autonomous system (AS),
that exploits all the IP addresses it owns to attack \ourmethod,
and that a certain number of honest nodes (100, 1000,
10000) are uniformly spread amongst the remaining currently active IP addresses.
Table~\ref{table:equiv-f-sampling-method}
shows the power $f$ of such an attacker, supposing that the attacker is the
Internet AS with the largest number of currently
active addresses (106 million in the dataset we used, spread over
5739 blocks).
This calculation shows that the hierarchical
sampling method reduces the power of the attacker down to 21\% when
only 1000 honest nodes run \ourmethod{}, where it would have been
above 99.99\% (i.e.~full control on the network) using uniform sampling.
Figure~\ref{fig:geolite_AS_B_n1000} shows the corresponding probability that \ourmethod will return Byzantine nodes by applying Equation~\ref{eq:solved} 
for the 100 biggest Internet ASes, assuming 1000 uniformly spread honest nodes.


\begin{figure*}[t]
	\centering
	\begin{subfigure}[t]{.24\textwidth}
		\includegraphics[width=\textwidth]{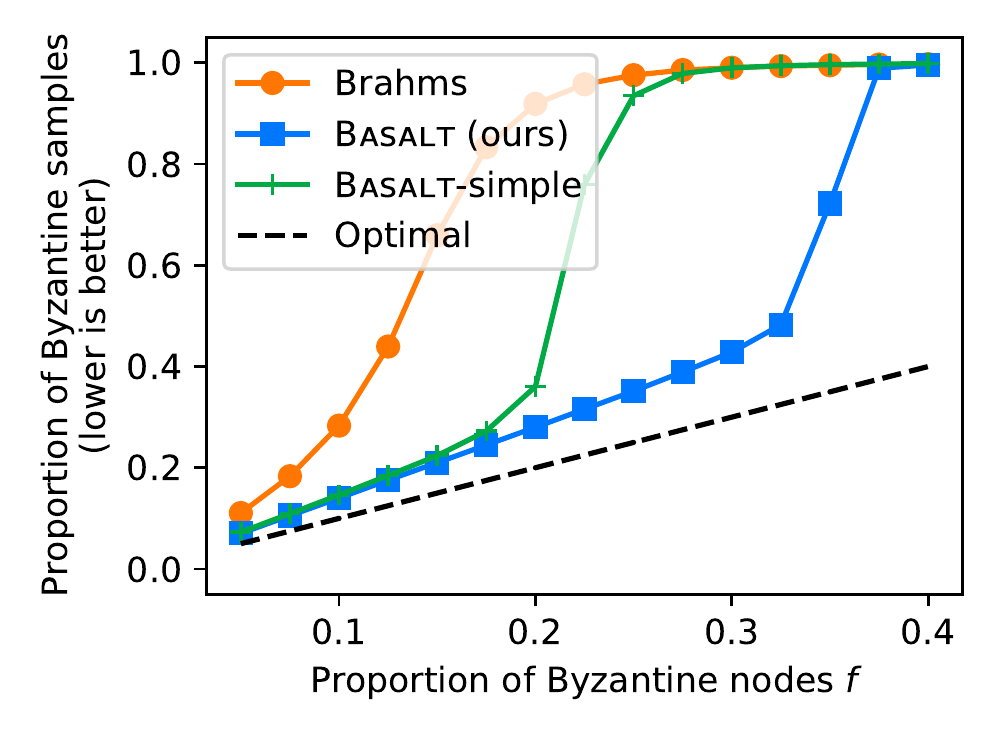}
		\caption{Varying the fraction $f$ of malicious nodes}
		\label{fig:success10k_tB}
	\end{subfigure}
	\begin{subfigure}[t]{.24\textwidth}
		\includegraphics[width=\textwidth]{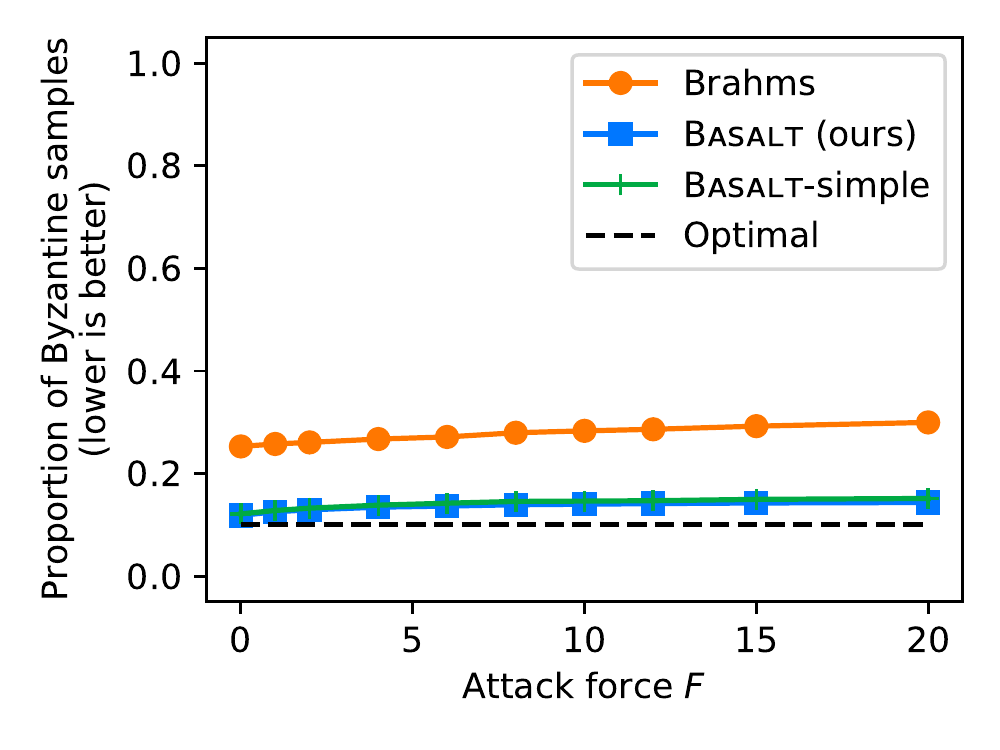}
		\caption{Varying the attack force $F$}
		\label{fig:success10k_fB}
	\end{subfigure}
	\begin{subfigure}[t]{.24\textwidth}
		\includegraphics[width=\textwidth]{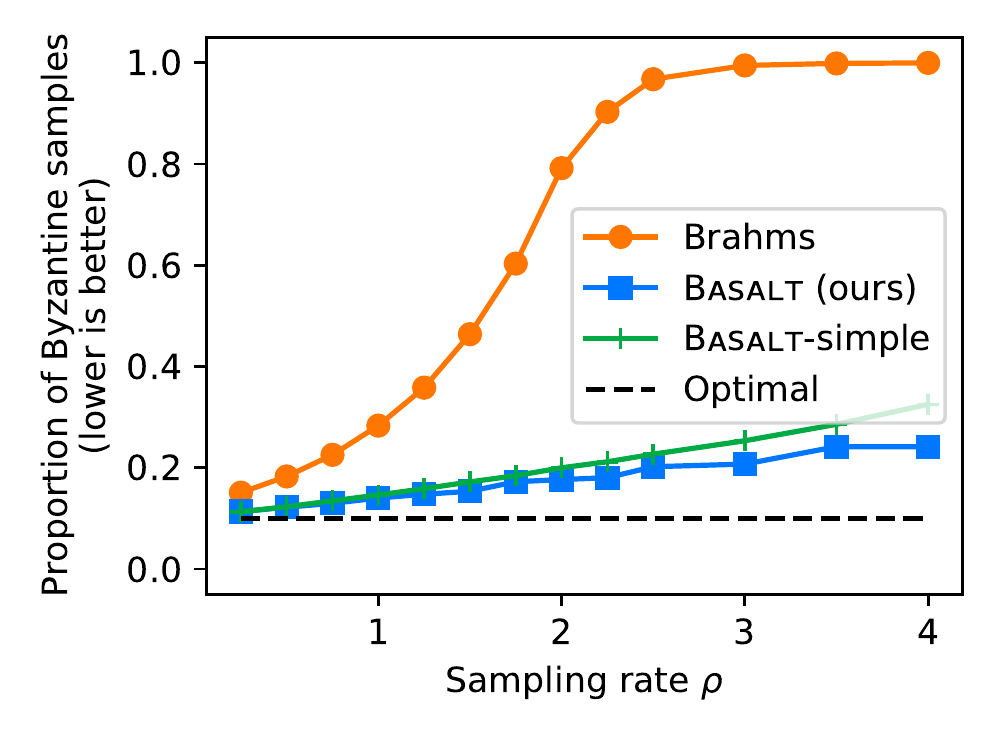}
		\caption{Varying the sampling rate $\rho$}
		\label{fig:success10k_RB}
	\end{subfigure}
	\begin{subfigure}[t]{.24\textwidth}
		\includegraphics[width=\textwidth]{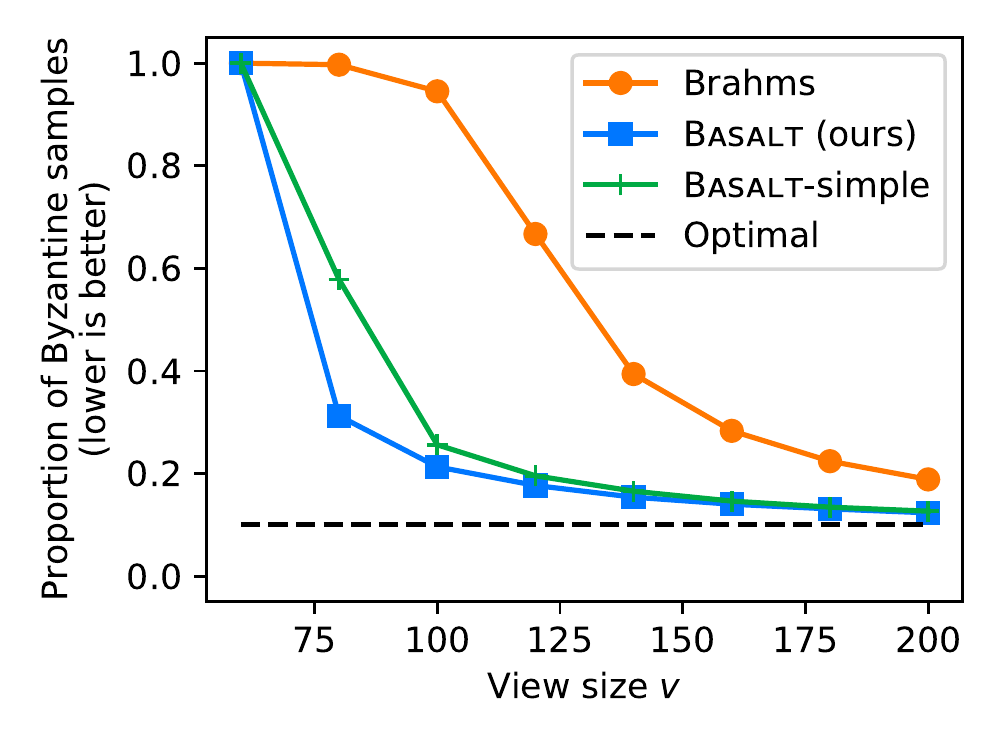}
		\caption{Varying the view size $v$}
		\label{fig:success10k_vB}
	\end{subfigure}
	\caption{Our algorithm (boxes, blue) consistently provides samples that contain
		fewer Byzantine nodes than our competitor, Brahms, in a variety of situations.
		A proportion of $1$ of Byzantine samples,
		as exhibited in Fig~\ref{fig:success10k_tB} for the highest values of $f$,
		corresponds to a situation where malicious nodes are able to cause a network partition.
		Results shown for a network size of 10000 nodes, with a base proportion $f=0.1$ of malicious nodes.
		Base values for other parameters are $v=160$, $\rho=1$, $F=10$.
		\ourm{} corresponds to the complete version of our algorithm,
		whereas \ourm{}-simple corresponds to Algorithm~\ref{alg:one} without the
		hardening mechanism (modifications of Algorithm~\ref{alg:two}).}
	\label{fig:success1k}
	\vspace{-1em}
\end{figure*}

\ft{Figure~\ref{fig:success1k}: Wondering is we should not add $v$ in the sub captions of (a) and (b).}

\section{Experimental Evaluation}
\label{sec:exp_simulation}

We complement our theoretical analysis with Monte Carlo simulations
that illustrate \ourmethod's dynamic behaviour.
In this section, we focus on simulating a permissioned system with a known fraction
of malicious nodes.
We do not simulate the IP address distribution and use the uniform ranking function.
As explained above,
our observations can be transposed to a permissionless setting, 
where the attacker's power defined by the hierarchical ranking function
plays the role of an equivalent fraction of malicious nodes.
We show that \ourmethod{} consistently produces samples with fewer malicious peers
than the state-of-the-art algorithms Brahms~\cite{bortnikov2009brahms} and SPS~\cite{jesi2010secure} over a wide range of scenarios.
We also show that \ourmethod{} converges faster on metrics
quantifying the random connectivity of the graph generated by the algorithm,
such as the clustering coefficient and mean path length.
These metrics are relevant for information dissemination and may thus have an influence
on the convergence time of epidemic agreement algorithms.

\subsection{Experimental Setting}
We evaluate the tested algorithms by simulating a system with $n$ nodes,
of which a fraction $f$ implement the malicious behaviour described in Section~\ref{sec:worst-case-attack}.
We do not simulate message loss or variable link latencies,
as our model parameter $F$ (the attack force) already integrates the possibility of message loss
(see Section~\ref{sec:worst-case-attack}),
and variable link latencies can also be modeled as losing messages that arrive after
a certain delay.
We do not simulate node churn, but consider instead an extreme scenario in which all nodes have just joined the system---this can be seen as an ultimate churn event, in which all nodes are replaced.
We vary the two parameters $v$, the view size, and $\rho$, the sampling rate, of the algorithm,
as well as the force of the attack, $F$.
We fix the exchange interval to $\tau=1$ ($1$ simulation time step).
Unless stated otherwise, we use $F=10$ and $\rho=1$.
All algorithms were implemented in a same simulation framework written in Rust,
totaling about 2500 lines of code~\footnote{\url{https://github.com/basalt-rps/basalt-sim}.}.

We compare \ourmethod{} (Algorithm~\ref{alg:one}) and its variant without the hardening mechanism (Algorithm~\ref{alg:two}, \ourmethod{}-simple) to two state-of-the-art competitors: Brahms~\cite{bortnikov2009brahms} and SPS~\cite{jesi2010secure}.

SPS was unable to function at all in the tested scenarios:
for instance for $n=1000$, $f=30\%$, and even with an attack force $F$ of $0$,
90\% of correct nodes become isolated in the network rapidly using SPS
and remain so during the whole simulation.
In contrast, both \ourmethod{} and Brahms were able to prevent all correct nodes from becoming isolated.
We have thus decided to exclude SPS from our comparison charts,
and concentrate on the comparison of \ourmethod{} against Brahms.

To compare Brahms and \ourmethod{} on similar grounds,
we add to the Brahms algorithm a mechanism that resets some of the hash functions regularly,
using the same round-robin strategy as \ourmethod{}.
Without such a mechanism Brahms would always return the same fixed set of samples,
limiting its usability as a random peer sampling algorithm.
As we will show just below, adding a reset rate to Brahms makes it less resilient to malicious nodes.%
\ft{Might we not get the criticism that we are weakening Brahms here? We should be careful how we present this.} 
In terms of communication overhead, Brahms and \ourmethod{} have the same cost.
Indeed, both algorithms send a set of peer identifiers of size $v$
when replying to a pull request.
For push requests, \ourmethod{} uses larger messages since Brahms does not send the view
with a push message, only the sending node's identifier,
whereas we send the whole view of size $v$.
However, supposing $v=200$ (the maximum in our
experiment) and node identifiers of size 4 bytes (such as IPv4 addresses),
the size of the communicated information is smaller than one MTU (maximum
transmission unit, i.e.~maximum size of a single packet, which is about 1500 bytes on the Internet),
thus the same number of Internet packets need to be sent by both algorithms.

\subsection{Proportion of Byzantine Samples}

In our first experiment, 
we measure the number of Byzantine nodes present in correct nodes' samples
on average after 200 simulation time steps.
For this experiment, we simulate a network of $n=10000$ nodes.
We fix base parameter values of $f=10\%$ of malicious nodes,
a sampling rate of $\rho=1$, a view size of $v=160$ and an attack force of $F=10$.
We then vary the parameters $f, \rho, v$ and $F$ individually.
Figure~\ref{fig:success1k} shows how this
proportion evolves for the three algorithms evaluated,
as one of the parameters $f,\rho,v$ and $F$ varies. 
\ft{`base' or `default'? `default' sounds clearer to me, but maybe that's just me.}%
\ft{Maybe some explanation/justifications for these 2 sets? From the results, parameter set 2 seems more challenging that parameter set 1. Could we give them names? (E.g. PS1 and PS2), or find a way to remind the reader which set is used in which Figure?}%
\ft{$v$ tend to vary quite a bit across the evaluation for $n=10000$ (e.g. $v=140,160,200$). It's better to stick to one default value unless there is a good reason to change.}

Plot~\ref{fig:success10k_tB} shows how the algorithms behave
when the proportion of Byzantine nodes in the system varies.
\Ourmethod{} provides close to optimal proportions of Byzantine samples 
even with many Byzantine nodes,
whereas Brahms fails to contain the attack in this domain.

Plot~\ref{fig:success10k_fB} shows the sensitivity
of the algorithms to the force of the attack $F$.%
\ft{The rationale behind figure groupings is unclear.}
These plots show that \ourmethod{} is almost insensitive to $F$,\ft{Why? Do we have an explanation?}
whereas Brahms shows an increasing proportion of Byzantine samples when $F$ increases.%
\ft{On the figures, I'd move `\ourmethod (ours)' to the last position, just before `optimal'.}

\begin{figure}[t]
	\centering
		\includegraphics[width=.7\linewidth]{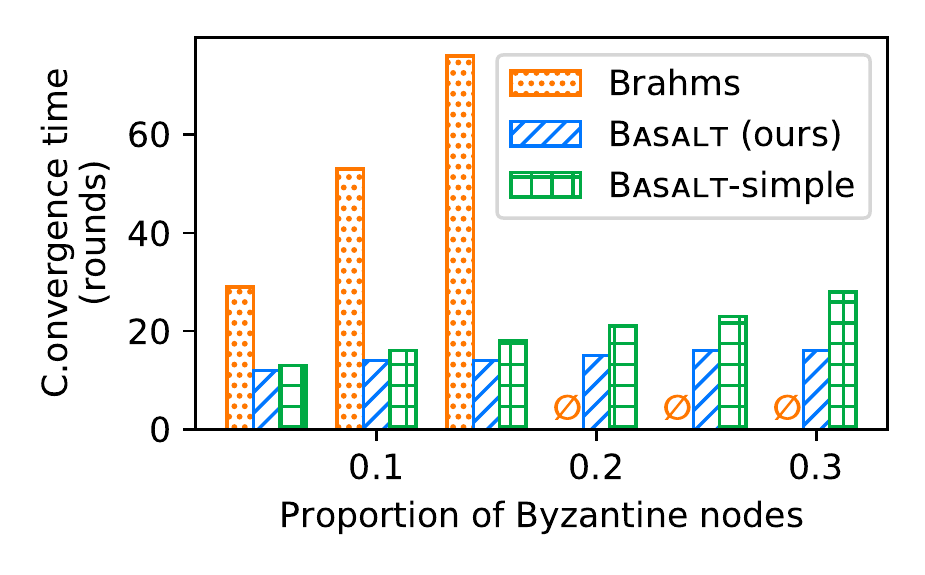}
		\vspace{-1em}
		\caption{Time to convergence within 25\% of optimal proportion of Byzantine samples, for $n=1000$, $v=100$ (on the right part, Brahms does not converge within experiment time)}
		\label{fig:bar_cv}
		\vspace{-1em}
\end{figure}

\begin{figure*}[t]
	\centering
	\begin{subfigure}[t]{.24\textwidth}
		\includegraphics[width=\textwidth]{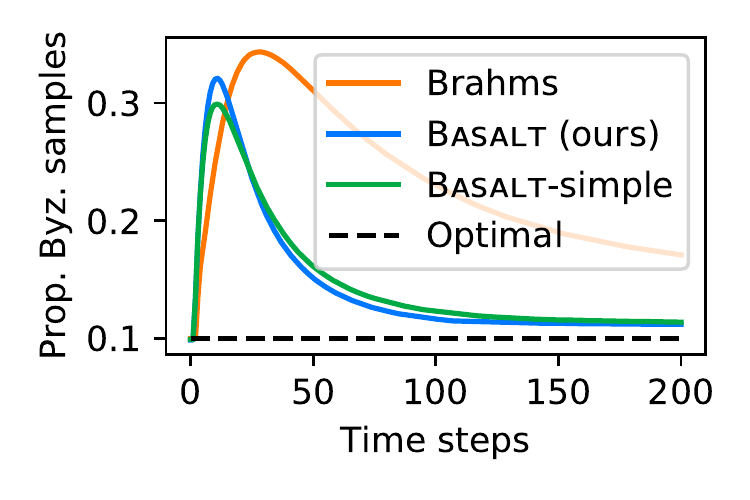}
	\end{subfigure}
	\begin{subfigure}[t]{.24\textwidth}
		\includegraphics[width=\textwidth]{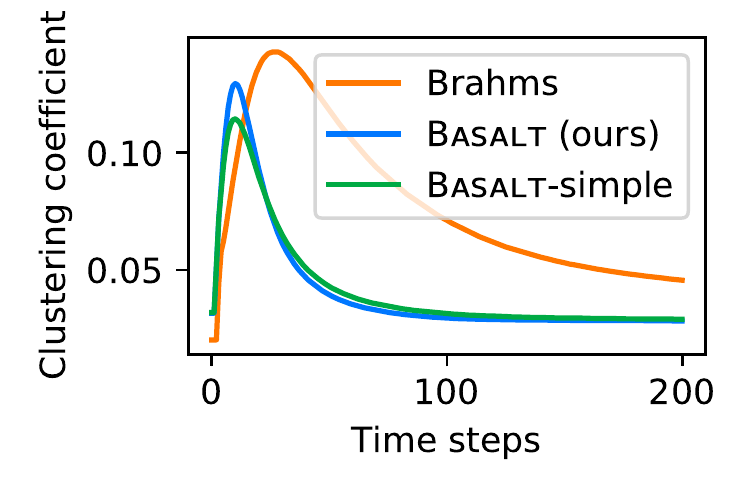}
	\end{subfigure}
	\begin{subfigure}[t]{.24\textwidth}
		\includegraphics[width=\textwidth]{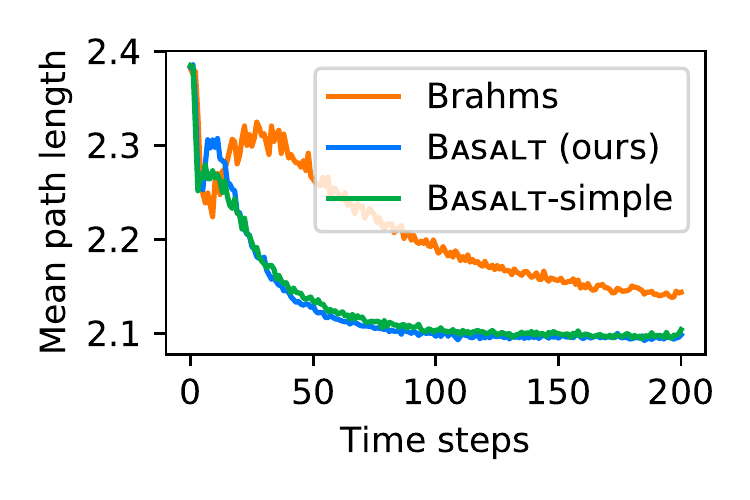}
	\end{subfigure}
	\begin{subfigure}[t]{.24\textwidth}
		\includegraphics[width=\textwidth]{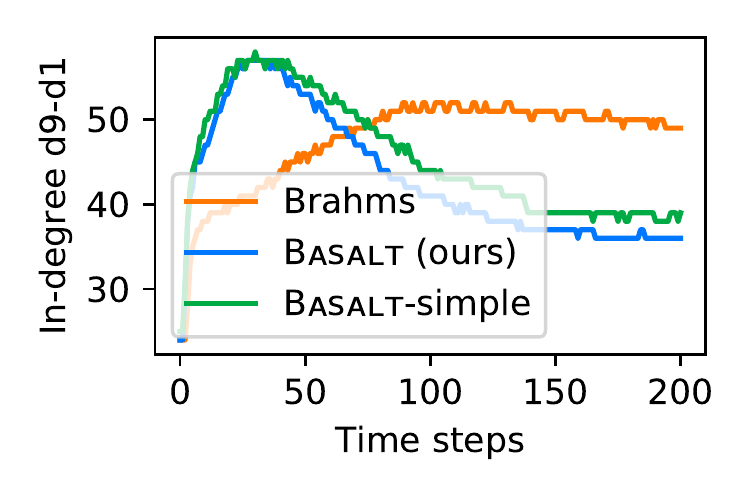}
	\end{subfigure}
	\vspace{-.5em}
	\caption{
		Algorithm convergence on several graph quality metrics, for $n=10000$, $f=10\%$, $F=1$, $\rho=0.5$, $v=160$.
		On all metrics, lower is better: we see that \ourmethod{} converges much more rapidly than Brahms.
	}
	\label{fig:convergence10k}
	\vspace{-1em}
\end{figure*}

Plot~\ref{fig:success10k_RB} shows how the algorithms
behave for various values of the sampling rate $\rho$.
For low values of $\rho$, both Brahms and \ourmethod{} are able to converge to
high quality samples, however such a setting does not provide much utility
as the algorithm is unable to frequently 
return new samples to the application.
Increasing the sampling rate $\rho$ results, however, in more disruption of the views,
where view slots have a higher risk of being reset before they converge to their target peer.\ft{I've moved up this sentence from Sec.~\ref{sec:compare_MR_max}, as we already discuss the effect of $\rho$ here.}
This disruption causes Brahms to collapse for higher values of $\rho$: the network becomes fully disconnected, and the views of correct nodes end up completely polluted by malicious peers.
This plot also shows how the hit-counter variant helps \ourmethod{}
attain better states when $\rho$ is high.%
\ft{The $y$ axis does not start at zero on many figures. It's a good practice to always start at zero (unless there are very good reasons not to), to make it easier to compare graphs, and avoid readers missing this important presentation point (as this makes differences look bigger).}

Plot~\ref{fig:success10k_vB} shows how the algorithms
behave for various view sizes. For small view sizes,
all algorithms are unable to keep the network in a connected state
and correct nodes all end up isolated.
The plots show that \ourmethod{} can keep the network connected
using smaller views than Brahms.

\subsection{Evaluating Convergence Speed}

\begin{figure}[t]
	\centering
		\includegraphics[width=.30\textwidth]{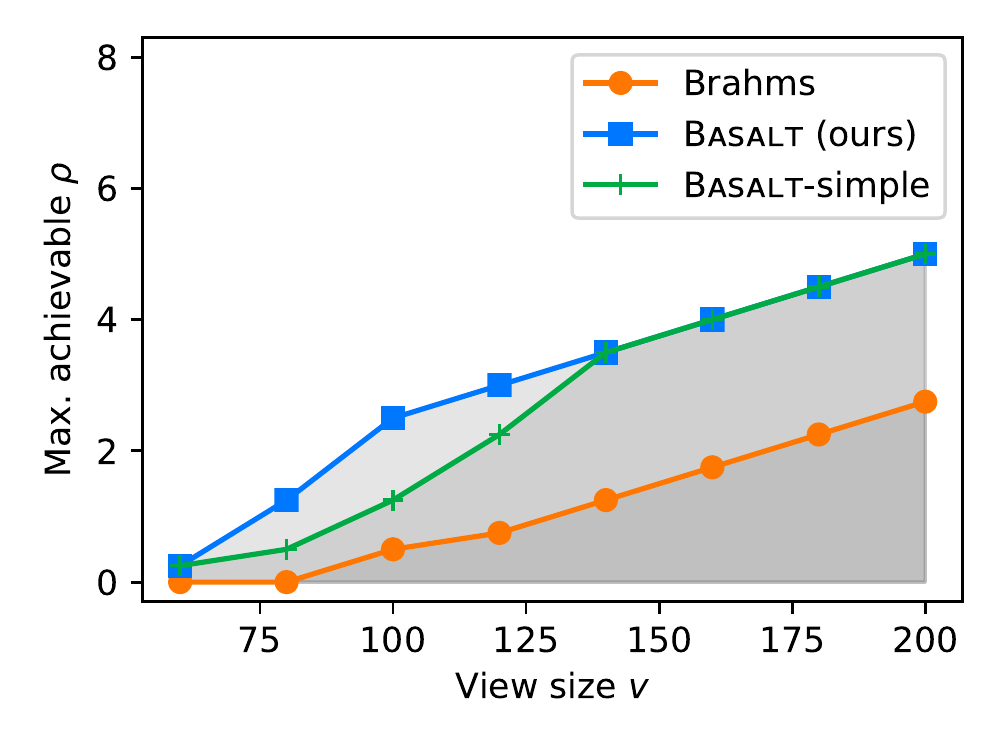}
		\vspace{-1em}
	\caption{
		Maximum achievable sampling rate $\rho$ (see Section~\ref{sec:compare_MR_max}) for 10000 nodes, $f=10\%$
	}
	\label{fig:compare_exp4_vR}
	\vspace{-1em}
\end{figure}

In this second experiment, we study the speed at which the algorithms converge
to good network states, where they provide samples with low proportions of malicious nodes.
Figure~\ref{fig:bar_cv} shows the time that Brahms and \ourmethod{} take to converge
to proportions of Byzantine samples that are within 25\% of the optimal proportion,
for $n=1000,v=100,F=10$ and $\rho=1$ and for varying proportions of Byzantine nodes in the network.
We show that the convergence time of \ourmethod{} remains low for up to 30\% of Byzantine nodes,%
\ft{Can we cite concrete numbers? E.g. so many rounds/time units? I'm also wondering if we should not use a real time unit, e.g. minutes or seconds, to better convey the simulation results. (So readers can make their own mind as to whether this can be useful/reasonable, otherwise remains a bit abstract.)}
whereas Brahms takes much longer to converge (starting at 20\% of Byzantine nodes,
it did not converge within the experiment's time).

Figure~\ref{fig:convergence10k} shows
the evolution of several metrics through time, starting with
the number of Byzantine nodes in the view,
in our experiment for $n=10000$, in a favorable situation with $f=10\%$, $\rho=0.5$ and $F=1$.
These plots show that \ourmethod{} converges much faster than Brahms
to a good network state (Brahms does not converge according to the previous
criterion within the time of the experiment).
The other plots show metrics for graph quality, where the algorithms exhibit
a similar convergence behaviour:
clustering coefficient, mean path length and the concentration of in-degrees measured by the difference
between the last and the first decile.
The clustering coefficient is computed by averaging the local clustering
coefficient of correct nodes in a graph where malicious nodes
are assumed to be all connected to one another.%
\ft{Do we need this assumption for the metric if we only consider correct nodes? If not, I'd drop the sentence.}
The mean path length is measured in a graph where malicious nodes
are assumed to have no connection in either direction, which models
the situation where they do not cooperate in transmitting information
between correct nodes.

\subsection{Node Isolation vs. Sampling Rate}
\label{sec:compare_MR_max}

We have seen earlier (Plot~\ref{fig:success10k_RB}) that both Brahms and \ourmethod{} are sensitive to increased sampling rates, and return more malicious samples
when the sampling rate, $\rho$, 
is high, with Brahms failing completely for too large
values of $\rho$.

To investigate this effect further, we run both algorithms for various values of $v$ and $\rho$,
and plot the maximum value of $\rho$ that can be used for a given $v$
without causing a network partition.
More precisely, a run for a given set of parameters $v,\rho$ is
successful if starting from half of the allocated simulation time, no
correct node is ever isolated by the malicious peers. Otherwise it is failed.
We plot the successful runs with highest values of $\rho$ for a given $v$.
The results of this experiment are shown in
Plot~\ref{fig:compare_exp4_vR} for $N=10000,f=10\%$ and $F=10$.
The areas delineated in Plot~\ref{fig:compare_exp4_vR}
correspond to the parameter sets that give successful runs.
Our results show that for similar view sizes, \ourmethod{} achieves much higher sampling rate
than Brahms, thus providing more utility to the application.

\section{Live Deployment}
\label{sec:exp_real_world}

We implemented \ourmethod{} in the AvalancheGo engine~\cite{avalanchego_url}, 
the main implementation of the AVA network~\cite{avalabs_url}
which uses the Avalanche consensus algorithm\footnote{Our code is publicly available at
\url{https://github.com/basalt-rps/avalanchego-basalt}.
Our implementation is forked from the official AvalancheGo repository~\cite{rocket2018snowflake}. Our changes are identified by ``Basalt RPS Authors''.}.
\commentFT{VLDB: Copied from USENIX Sec. response}%
We picked AVA, as it is the main cryptocurrency network that uses an epidemic, sampling-based
consensus, which is the target use case of \ourmethod.
Our implementation, a 500-lines patch to the Go source code
of AvalancheGo, replaces peer sampling based on stake in a
proof-of-stake system by peer sampling based on \ourmethod,
including the hierarchical ranking function  described in Section~\ref{sec:sybil_biased_sampling}.

Our implementation integrates seamlessly with the AVA protocol\df{Why   should it?} and is fully compatible with the existing network.
Our implementation supports managing current outgoing connections according to the \ourmethod{} algorithm,
instead of keeping connections open to all reachable network nodes as done by the original AvalancheGo implementation%
\footnote{Unfortunately, we had to disable this behaviour as it led to too many connection attempts
and some nodes appeared to have banned our IP addresses as a consequence.
A simple modification allows our code to never close connections intentionally:
the view maintained by \ourmethod{} is only used to sample peers for the Avalanche consensus algorithm,
and connections are kept in the background to nodes that have been removed from the view.}.

To show that \ourmethod{} can be applied as a sampling method that reduces the risk of an
\institutional attack, we ran a 10-hour experiment where we launched 100 ``adversarial''
Avalanche nodes
on the public AVA network (corresponding to about 20\% of total active nodes)
in an attempt to bias sampling in their favor using a Sybil attack against one of our nodes.
The nodes we launched all had IP addresses located in the same /24 prefix,
owned by our research institution.
Samples were measured at witness nodes running the \ourmethod{} sampling algorithm,
as well as the non-hierarchical variant of \ourmethod{} and a sampling algorithm based
on full network knowledge.
Results shown in Table~\ref{table:basalt_ava_sampling} show that using \ourmethod{}, 
the probability of sampling one of our adversarial nodes is brought to about 1\%,
meaning that the influence of our nodes in the network is extremely limited.

\begin{figure}[t]
	\centering
		\includegraphics[width=.95\linewidth]{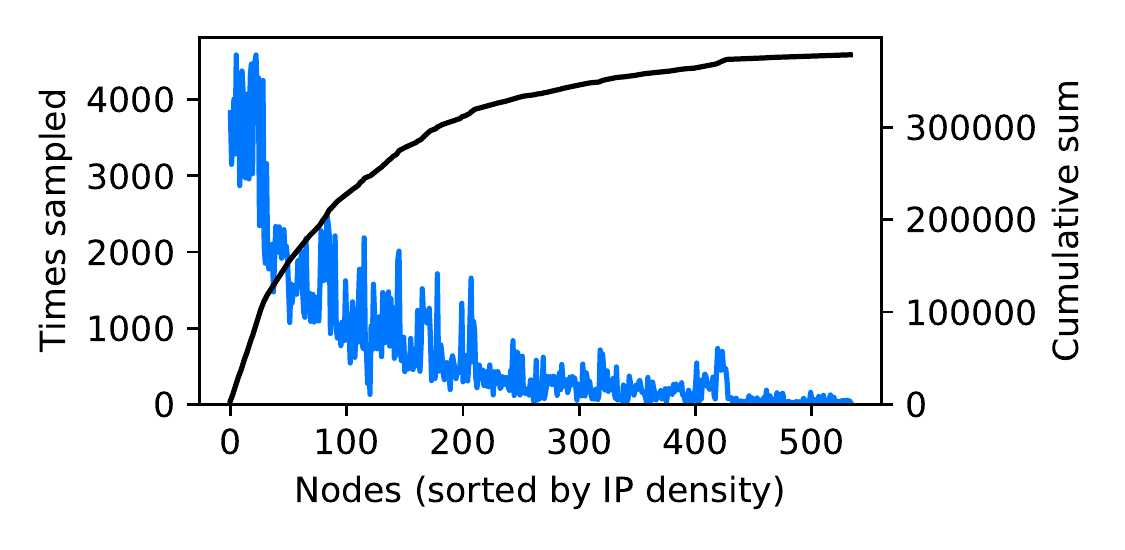}
		\caption{Behaviour of AvalancheGo modified with \ourmethod{} running on the public AVA network, 5-hour experiment.
			On the left, nodes that are alone in their IP prefix are sampled the most frequently.
			On the right, nodes that belong to IP prefixes where other network nodes are present
			are sampled less often.}
		\label{fig:basalt_ava}
		\vspace{-1em}
\end{figure}

To show the wider benefit of \ourmethod{}, we plot in Figure~\ref{fig:basalt_ava}
the number of times the various nodes of the AVA network were sampled in the
experiment. Sorting nodes by a density metric which counts the number of other
nodes in the same /8, /16 and /24 prefix reveals that nodes which are isolated in their prefix
(on the left of the graph) are sampled more often than nodes which share their
IP prefixes with other nodes (on the right). However all network nodes
have a chance of being sampled, and no single node is sampled exceedingly more
often than others. This stands in contrast with Proof-of-Stake-based sampling,
where sampling frequency is proportional to the stake invested,
a mechanism that gives disproportionate power to rich nodes and totally excludes
nodes that are not able to invest any stake in the network.

\begin{table}[t]
	\small
	\begin{center}
		\caption{Observed proportion of samples that are nodes controlled by the adversary
		in our live experiment (see Section~\ref{sec:exp_real_world})}
		\label{table:basalt_ava_sampling}
    \begin{tabular}{|l|c|}
		  \hline
		\textbf{Algorithm} & \textbf{Adversary samples}  \\
		\hline
		Full knowledge uniform sampling & 18.4\% \\
		\hline
		\ourmethod{}-uniform & 17.5\% \\
		\hline
		\ourmethod{} (hierarchical) & \textbf{1.13\%} \\
		\hhline{|=|=|}
		True proportion of Byzantine nodes & 18.8\% \\
			\hline
		\end{tabular}
	\end{center}
	\vspace{-1em}
\end{table}

\section{Discussion and Related Work}
\label{sec:related_work}
Random peer sampling in non-adversarial settings is a well-studied
problem~\cite{jelasity2007gps,voulgaris2005cyclon}.
Surprisingly, very few works have sought to develop Byzantine-tolerant
RPS protocols.

State-of-the art methods such as
Brahms~\cite{bortnikov2009brahms} and Secure Peer
Sampling~\cite{jesi2010secure} (SPS)
are based on a classical RPS
algorithm, to which is adjoined a mechanism that tries to correct for
the over-representation of malicious nodes.
In Brahms, the view is not updated if a peer has received more than a
certain number of push messages in a given time slot.  Albeit vaguely
similar to our hit counter mechanism, Brahms' approach can only work
if we assume that malicious nodes have limited total firing power and
must therefore target their attack on a specific victim node.
Otherwise, they would be able to simply flood the whole network with
many pushes and halt the peer sampling algorithm completely.
In SPS,
nodes try to build some statistical knowledge on node behaviour;
however, this mechanism is unable to cope with attacks where malicious
nodes send so many messages that correct nodes do not have the time to
gather sufficient statistics to block them before becoming isolated.
Our protocol, on the other hand can effectively handle these attacks.
Moreover, the majority of these systems
do not address risks that exist in real-world networks, such as Sybil
attacks. To our knowledge, the only exception is HAPS~\cite{amaury}, which is designed specifically to handle Sybil attacks. 
\commentFT{VLDB: Copied from USENIX Sec. response}%
HAPS, however, only addresses Sybil attacks in which attackers are
concentrated in a few IP blocks ("\institutional attacks"), by using random walks on a
carefully crafted probabilistic tree. Due to its design, it is not immediately
clear how HAPS could be extended to counter attackers that are spread out,
which \ourmethod does thanks to its stubborn chaotic search.

\label{sec:network-attack}
Recent works on blockchains have also brought to lighten the risk of
attacks at a more fundamental level than those described in Section~\ref{sec:sybil_attack_def}.
Network adversaries are
malicious entities that gain control of part of the routing infrastructure
(internet autonomous systems, or ASes),
in which case they can intercept and modify all the traffic that they are
routing, or attack the routing algorithm itself by
advertising Internet prefixes that they do not own, thus attracting
traffic that should have gone through another path, a so-called
\emph{BGP hijack}~\cite{hijackbtc2017}.

Note that BGP hijacking attacks are necessarily limited to one or a few IP prefixes,
as large-scale routing attacks would likely bring down large parts of the Internet
and would be noticed immediately.
By spreading connections over a variety of IP prefixes through its $\mathrm{rank}$ function,
\ourmethod{} builds intrinsic resilience to these attacks as at most only a small fraction
of nodes' neighbors will be located in hijacked prefixes.
In this way, the global \ourmethod{} network is not at risk of being taken down
or manipulated by a malicious entity.

However, network attacks might also be used to target specific nodes,
to remove them from the global network and make them believe
false information about the network's state (an Eclipse attack).
Defenses have been proposed against Eclipse attacks at the network level: for instance,
the SABRE network~\cite{sabre} proposes to use additional communication channels,
in the form of a network of specialized nodes that are all connected to one another
using dedicated channels,
and that are located close to end-users
so that they can provide a safe service directly to them even in the case
of a hijack.

In the case of a blockchain, where the most crucial property to
guarantee safety is that all nodes are made aware of new blocks
rapidly, the SABRE method is able to help by providing reliable block
delivery.  For sampling-based methods that use \ourmethod{}, SABRE
could provide a security mechanism at the application layer to enable
detection of network attacks and stop all activity in case they
happen, for instance by detecting a discrepancy between a node's local
state and the state of SABRE nodes.  This mechanism however cannot be
used to allow eclipsed nodes to make progress in such a situation, as
it does not provide the secure random peer sampling service itself.
Finding mechanisms to allow nodes that are eclipsed by a network
attack to continue functioning normally when running a sampling-based
algorithm is, to the best of our knowledge, still an open problem.\ft{I have merged the discussion with the related work. To be discussed.}

\commentFT{VLDB: Copied from USENIX Sec. response}%
Finally, one could argue that it will be hard to bootstrap a \ourmethod network
  containing enough nodes to effectively counter botnet attacks.  We note that
  this problem is exactly the same as in PoW-based cryptocurrencies, as an
  attacker that gains $>50\%$ of the network's hashing power can overturn the
  network in their favor (which is easy to do for smaller cryptocurrencies that
  don't have a lot of hashing power allocated to them).  A PoW-based
  cryptocurrency network is secured by members investing in providing lots of
  hashing power, as is the case e.g. for Bitcoin, in order to make a $>50\%$
  attack so costly that it is impossible in practice (or simply not worth it
  compared to the value of the cryptocurrency that could be stolen).  A
  \ourmethod-based cryptocurrency is similarly secured by participants investing in
  running as many nodes as possible from many different IP prefixes, which they
  have an incentive to do in order to keep the system safe. Moreover, \ourmethod
  has the advantage that this investment does not require the waste of
  tremendous quantities of energy.

\section{Conclusion}
\label{sec:conclusion}
\balance

We have presented a new algorithm for Byzantine-tolerant random peer sampling
on the Internet that uses biased sampling to prevent Sybil attacks.
Such an algorithm can be used to implement sampling-based consensus algorithms
such as Avalanche. Contrary to sampling algorithms based on Proof-of-Stake,
such as those currently in use on the AVA network, \ourmethod{} allows the
network to be truly open by allowing any Internet user to join the consensus
without having to own any cryptocurrency tokens.
We expect that in the future the line of research around Byzantine fault-tolerant algorithms
based on epidemics will continue to see new developments motivated by gains in performance,
and thus we believe that we have brought an important contribution
to making such methods applicable in large-scale open networks.




\appendix

\section{$C(\correctSet,\byzSet)$ in a Botnet attack}
\label{sec:botnet-attack}
\newcommand{\select}{\mathsf{selected}}
The probability $C(\correctSet,\byzSet)$ depends on the distribution of correct and Byzantine identifiers across the three levels of blocks used in Equation~\ref{eq:rank:def}. We fix one node $p$ selected randomly amongst $\correctSet\cup\byzSet$, and write $\select(p)$ the event that $p$ is selected by the ranking function $\rank{\cdot}{S}$:
\begin{align}
  \select(p) \equiv \Big(p = \mathrm{argmin}_{q \in \correctSet\cup\byzSet} \rank{q}{S}\Big).
\end{align}
With this notation we have $C(\correctSet,\byzSet)=\Pr\big(\,p\in\correctSet\,|\,\select(p)\,\big)$.

In our model, a botnet attack corresponds to the (ideal) case in which Byzantine and honest nodes follow the same distribution across IP blocks. As a result, they are indistinguishable from the point of view of $\rank{\cdot}{S}$, which means here that the events $p\in\correctSet$ and $\select(p)$ are independent. This independence implies that
\begin{align}
  C(\correctSet,\byzSet)=&\Pr\big(p\in\correctSet|\select(p)\big)\nonumber\\
  =&\Pr\big(p\in\correctSet\big)\nonumber\\
  =&\frac{|\correctSet|}
     {|\correctSet|+|\byzSet|}.
\end{align}

\section{Deriving Equation~(\ref{eqn:convergence-isol-state-result})} 
\label{sec:apx-convergence-isol-state}

Based on the result from the coupon collector's problem,
the expected number of uniformly distributed (non-distinct) correct peer identifiers
that must be received in order to learn $\Delta c$ new \textbf{distinct} correct peer identifiers
amongst $Q$, when $c_0$ are already known, is:
\begin{dmath}
	\frac{Q}{Q-c_0}+\frac{Q}{Q-c_0-1}+\cdots+\frac{Q}{Q-c_0-\Delta c+1}
	\label{eqn:q-qc0-deltac}
\end{dmath}
The number of uniformly distributed peer identifiers received between the
two resets is at least the following expression:
\begin{dmath}
	\frac{k}{\rho} \frac{v}{\tau} \frac{c_0}{f n + c_0} (1-f)
	\label{eqn:k-rho-v-tau-c0-f}
\end{dmath}

where $\frac{k}{\rho}$ is the duration of the considered time slice,
$v$ is the number of peer identifiers exchanged at each exchange step,
$\tau$ is the time between two exchange steps, $\frac{c_0}{f n + c_0}$ is the
probability that the exchange was conducted with a correct peer,
and $(1-f)$ is the probability that each of the peers of the returned view
is correct.

We bound the value of $(\ref{eqn:q-qc0-deltac})$ as follows:
\begin{dmath}
		(\ref{eqn:q-qc0-deltac}) \le \Delta c \frac{Q}{Q-c_0-\Delta c}
\end{dmath}

Moreover, we have $(\ref{eqn:q-qc0-deltac}) \ge (\ref{eqn:k-rho-v-tau-c0-f})$.
Thus:
\begin{dmath*}
	\Delta c \frac{Q}{Q-c_0-\Delta c}
		\ge 
	\frac{k}{\rho} \frac{v}{\tau} \frac{c_0}{f n + c_0} (1-f)
\end{dmath*}
thus
\begin{dmath*}
	\Delta c Q \tau \rho (f n + c_0) \ge k v c_0 (Q - c_0 - \Delta c)(1-f)
\end{dmath*}
thus
\begin{dmath*}
	\Delta c \ge \frac{k v c_0 (1-f)(Q-c_0)}{Q \tau \rho (f n + c_0) + k v c_0 (1-f)}
\end{dmath*}
which is the result of Equation~
	$(\ref{eqn:convergence-isol-state-result})$.

\clearpage

\bibliographystyle{ACM-Reference-Format}
\bibliography{bibliography}

\end{document}
